# A Framework for Detecting Event related Sentiments of a Community

By

Muhammad Aslam

Jarwar

**Department of Computer Science**
**Quaid-i-Azam University**
**Islamabad, Pakistan**
**April 20, 2019**

Dedicated to

***H & H (Hasnain & Hibbah) My Kids***

# Declaration

I hereby declare that this dissertation is the presentation of my original research work. Wherever contributions of others are involved, every effort is made to indicate this clearly with due reference to the literature and acknowledgment of collaborative research and discussions.

This work was done under the guidance of Dr. Rabeeh Ayaz Abbasi, Department of Computer Sciences, Quaid-i-Azam University, Islamabad.

Date: April 20, 2019

Muhammad Aslam Jarwar

# Abstract

Social media has revolutionized human communication and styles of interaction. Due to its easiness and effective medium, people share and exchange information, carry out discussion on various events, and express their opinions. For effective policy making and understanding the response of a community on different events, we need to monitor and analyze the social media. In social media, there are some users who are more influential, for example, a famous politician may have more influence than a common person. These influential users belong to specific communities. The main object of this research is to know the sentiments of a specific community on various events.

For detecting the event based sentiments of a community we propose a generic framework. Our framework identifies the users of a specific community on twitter. After identifying the users of a community, we fetch their tweets and identify tweets belonging to specific events. The event based tweets are pre-processed. Pre-processed tweets are then analyzed for detecting sentiments of a community for specific events. Qualitative and quantitative evaluation confirms the effectiveness and usefulness of our proposed framework.

# Acknowledgment

First of all, I am thankful to the most Merciful and the most gracious ALLAH for giving me the courage and strength to complete this challenging job. I feel immense gratitude for my supervisor, Dr. Rabeeh Ayaz Abbasi, whose guidance and consistent encouragement made it possible for me to complete this research.

Also Dr. M. Afzal Bhatti, Dr. Shuaib Kareem, Dr. Khalid Saleem and department faculty deserve extraordinary credit for their splendid coaching during the course work and especially in research, which helped me at different stages of this study.

I show great appreciation to my class mates and friends who directly or indirectly helped or encouraged me at any point. They provided tremendous rapture and delight, which calmed me whenever I felt dejected. I am also thankful to my whole family for their consistent encouragement, support and motivation to carry on this work. In a nut shell, this achievement was not possible for me without the support, encouragement and prayers of my family, friends and all others I forget to mention here.

Thank you,

*Muhammad Aslam Jarwar*

# Contents

# List of Tables

# List of Figures

# Chapter 1

# Introduction

This chapter introduces the proposed research work and discusses the motivation behind this research. The structure of chapter is as follows: Section 1.1 discusses the background. Section 1.2 elaborates the motivation behind the study. Section 1.3 lists the hypothesis of our research. Section 1.4 highlights the challenges of the addressed problem.

## 1.1 Background

In the era of Web 1.0, there were very few websites mostly having static pages. Contents were created by the authors and domain experts of these websites in structured and manageable format. In Web 1.0 users did not create, share and exchange information. They acted only as the consumers of information. Due to limited creation of information and its structured and manageable format, it was easy to analyze and filter information.

The birth of Web 2.0 gave the boom to creation of many dynamic websites and services. In Web 2.0 the most popular and successful websites, services and application belonged to social media services and networks. Among these, most popular are Facebook[1], Twitter[2], LinkedIn[3], Pinterest[4], and Google Plus[5] [1].

Social media websites and applications revolutionized human communication and style of interaction. In this style of interaction humans create, share and exchange information easily and effectively.

Due to ease and effective medium of communication and sharing of information, opinions and emotions; people communicate and share their problems directly to their representatives in government and parliament (e.g. Pakistan Tehreek-e-Insaf MNA Asad Umar vs Ahsan as in Figure 1.1) and also give their opinions and show their emotions on social problems, social events, political movements and government policies. Active participation of a large number of users results in abundance of information and most of this information is unstructured and unmanageable. The huge amount of information in social media creates the problem of "Social media information overload". Social media information overload and diversity of information creates difficulties and challenges in information processing, presentation and analyzes.

## 1.2 Motivation

In social media, the information which is created, shared and exchanged has importance for public, news agencies, marketing companies, governments, oppositions and political parties; because this information contains public opinions, emotions, and sentiments. News agencies select subject of talk shows and the trends of news as per opinion and emotions of public in social media. For example in Pakistan International Airlines two lawmaker personalities were forced to leave the aeroplane on 16th September 2014. This story broke through one of the most popular online social media service Twitter .

[1] https://www.facebook.com/
[2] https://twitter.com
[3] https://www.linkedin.com
[4] https://www.pinterest.c

om/
[5]https://plus.google.com/

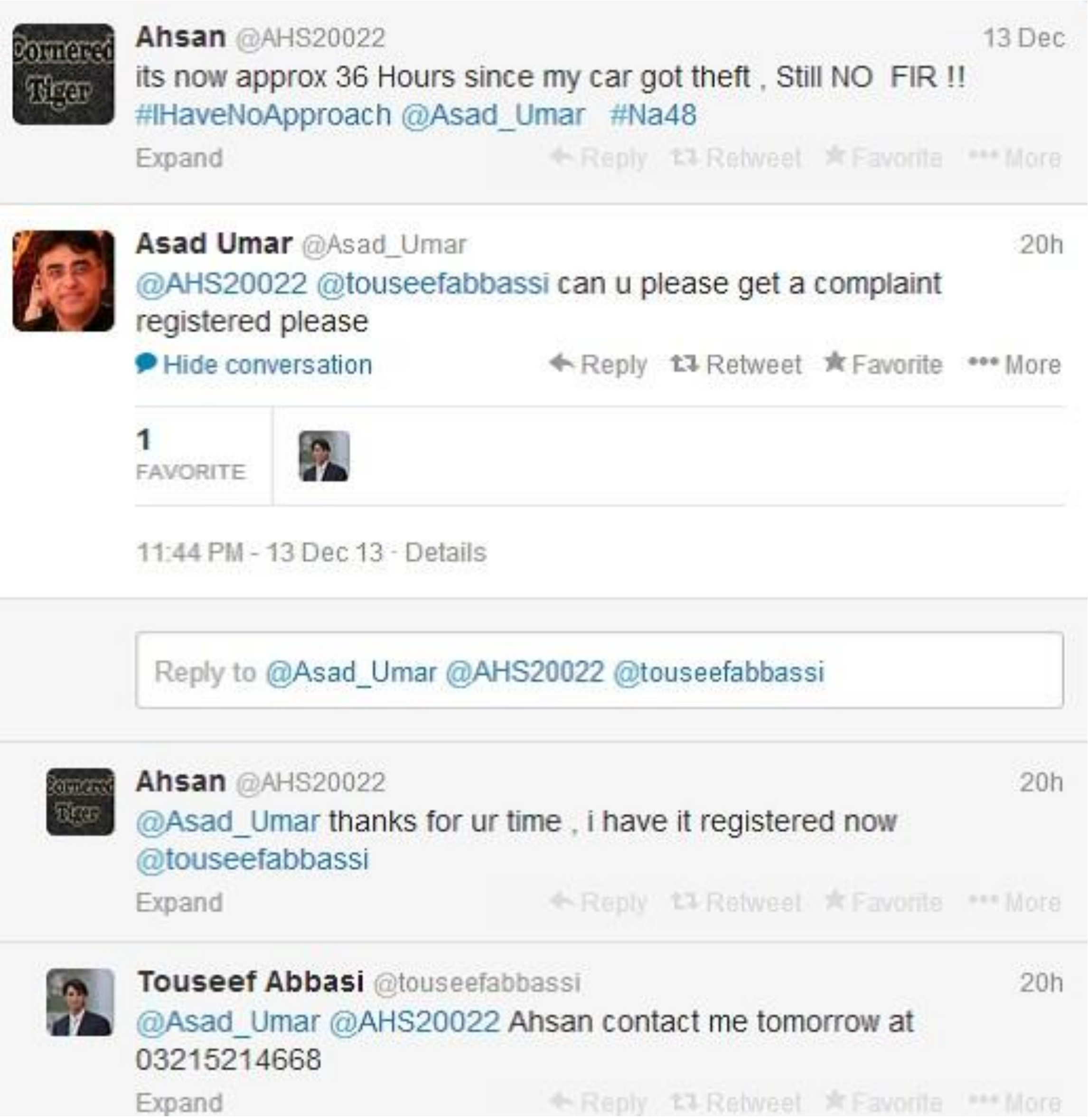

Figure 1.1: This figure shows the conversation between Mr. Ahsan, a common person and Mr. Asad Umar, MNA of a political party "Pakistan Tehrik-iInsaf". Mr. Ahsan's car was stolen, he was unable to register a FIR (First information report) at the relevant police station. He shared his problem with the MNA of his constituency Mr. Asad Umar. Mr. Asad Umar directly response and helped him. [a]

[a] https://twitter.com/Asad_Umar

Marketing companies target communities to get valuable information about their living styles, and get feedback of the products these communities use in their daily life. This is possible because these communities share their experiences about the brands they use in their daily life on social media. For example marketing companies monitor user comments, opinion, emotion and sentiments to see what consumers think of your product or business directly. By monitoring social media, marketing companies set their goals and define their strategies accordingly.

A government may also be able to get benefits from social media while making its policies and decisions about the country and its public, as users on social media discuss and give their opinions about the government policies, decisions and its governance among their friends, colleagues and general public. Through the effective monitoring and analysis of social media posts, government may make their policies and execute government decision in a better way [2].

The oppositions in democratic countries may perform their role of vigilance related to government and its policies by taking input from social media posts, because users of social media frequently share their problems and stories of social injustice, which contain the user's opinions, emotions and sentiments.

The major political parties in Pakistan have established their social media teams, which share party agenda, speeches of party leaders on social media, and also forward messages to targeted communities. The main reasons behind the establishment of social media teams, is the recognition of the importance of public opinion and sentiments on the social media. On 14th August 2014, when the two political parties Pakistan Tehrik-e-Insaf (PTI) and Pakistan Awami Tehrik (PAT) started their Azadi and Inqlab March respectively, the social media flooded with the diversity of public opinions and emotions. Some groups talked in the favor and others opposed and while some remained neutral.

In real life, there are many communities like lawyers, politicians, doctors, and researchers. Now a days most of these communities are aware of the importance of social media and use social media services in their daily lives. Among these communities, the journalist community actively participates in discussions on social media like twitter, and express their opinions about the events occurring in their surroundings. Journalists and media have strong influence on government policies and they effect the mindset of the public, which also impact the election results. Journalists use social media services [3] [4] together the news about the major events. Mostly news of major events comes from microblogging and social media services, e.g. revolutions of the Arab spring, death of former British Prime Minister Margaret Thatcher and explosions at the Boston marathon[6].

From the above discussions, we conclude that in social media collectively user opinions and sentiments have very importance role. Millions of users share billions of posts on popular social media services, mostly in the form of microblogging. The most popular microblogging service is "twitter" used by millions of users. Therefore analyzing the opinions and sentiments of communities may benefit in knowing the public mood in a quantitative as well as qualitative way. By using these results; news agencies, marketing companies, governments, oppositions and political parties may perform better than before.

## 1.3 Hypotheses

In Section 1.2 we discussed the motivation behind our study. On the basis of the motivation, we hypothesize that:

*"1. Communities tweet on the events occurred in their surroundings."*

*"2. Communities express their opinions in the tweets on the events occurred in their surroundings."*

[6]http://www.theguardian.com/technology/2013/apr/23/twitter-first-source-investment-news

## 1.4 Challenges

There are many challenges in verifying these hypotheses. For our study we require a dataset containing the micro-posts of a targeted community. For collecting the micro-posts of a targeted community, we require a microblogging social service, like twitter and user-ids of the members of the targeted community. The members of the targeted community should not be favorite of someone and must present the entire community. Hand picking the members of a community would introduce biasness in the results of the research. To avoid any biasness we have to use a sampling method which represents the entire community.

After identifying the members of a community we need to collect their micro-posts (i.e. tweets). Twitter imposes rate limit on collecting tweets. Currently twitter API returns 3200 most recent tweets posted by a user. In the interval of 15 minutes/window [7] twitter API allow 15 requests per window. So it is difficult and time consuming to collect maximum number of tweets in a short time period. For collecting tweets, we may have to write a good crawler. It is also a challenging task to process the large collection of tweets for the feature extraction. Therefore we have to define a mechanism to divide large collection of tweets into relevant chunks or separate tweets according to focused topics or events. We must know the metadata of focused topic so that we can easily separate our data-set into chunks or relevant topics.

The tweets (micro-posts) include many meta-data features; such as slang words, abbreviations, hash-tags, URLs, mentions and user-ids and tweets also have short length, without any grammatical rules and highly irregular structure. So getting our desired features from tweets is also a challenging and interesting task.

Our data-set may contain the data in Urdu and Roman Urdu languages, and we have to extract features from that data as well. Although most of the open source

[7] https://dev.twitter.com/rest/reference/get/statuses/user_timeline Last Access on 2015-02-21 23:58

and best classifiers support English language. These open source English language classifiers classify the text on the basis of entire documents and bags of positive and negative words. We need a classifier that detects sentiments from short length of text (i.e. tweets and micro-blogs) on the basis of syntactic and semantic structure of language. The last challenge in our study to quantitatively and qualitatively analyze the results.

## 1.5 Summary

In this chapter we have discussed background, motivation, hypotheses and many challenges related to our study. Further we focused on the importance of Web 2.0 and social media among different communities and in real life.

# Chapter 2

# Related Work

This chapter discusses literature review of our study. The structure of chapter is as follows: Section 2.1 introduces definition of microblogging. Section 2.2 gives overview of twitter and terminologies. Sections 2.4, 2.5 and 2.6 provides definitions of opinion, sentiment and emotions. Section 2.7 discusses the relation among opinions, sentiments and emotions. Section 2.8 gives an overview of sentiment analysis and its techniques. Section 2.10 elaborates snowball sampling and its significance in our study. Section 2.11 highlights different evaluation methodologies and its usage.

## 2.1 Microblogging

According to oxford dictionary the meaning of Micro-blog is “a social media site to which a user makes short, frequent posts”. A Micro-blog is very small than a normal blog. Mostly a single Micro-blog contains one sentence with or without a link or image or a video. Microblogging is popular for informal communication [5] and sharing of ideas on social media networks. Micro-blogs can be used in many different ways. Some people share their daily activities and update their status in form of a micro-blogs and others give their point of view about the topic. Many social networking websites support microblogging. Popular examples of microblogging websites are Twitter and Weibo[1]. Weibo is mainly popular among Chinese community, whereas Twitter is popular worldwide [6, 7]. Due to popularity of twitter as a social media microblogging service, we use twitter posts in this research.

## 2.2 Twitter Overview

Twitter was launched officially on 13th July 2006 [8]. It facilitates its users to communicate in a real time and create, send, receive and read posts known as "Tweets". The length of tweet is limited to 140 characters [9] and averaging eleven words per tweet [10]. The restriction of 140 characters supports the argument that twitter is a microblogging service. Twitter is also popular in academic researchers [9], Because most of the tweets are publically available and are accessible through twitter API. Users who follow other users for receiving their tweets are called followers and to whom they follow is called followee. In twitter, it is not compulsory that users follow each other reciprocally. In twitter only 22.1% users follow each other [11]. Different activities are performed by twitter users, such as to post a tweet publicly or specially to a user by mentioning his address as "@userid" (see in figure 2.2) , read a tweet, and forward a tweet known as "Retweet". The retweet mechanism of twitter gives the strength to users to spread a tweet to many users who are not followers of the original user who created the tweet. Due to its specific structure and features, twitter has emerged as a new medium of communication and a channel of rapidly spreading information [11, 12].

Twitter users use hash symbol "#" followed by word called as "hashtag" (see in figure 2.2 in their posts to categorize posts or follow posts related to a specific topic. Sometimes users overlap topics by using hashtag not in the context of the topic [13]. Hashtags further help users in searching posts.

Hashtags, simple words, and phrases which are used by many users in their tweets are tracked by twitter for detecting trending topics. Twitter by default shows ten trending topics on every user's home page as seen in figure 2.1. The user may configure trending topics at regional or global level. The exact mechanism behind the trending topics is unknown [11].

[1] http://weibo.com/login.php?lang=en-us

Twitter users show their feelings in tweets using emoticon. Emoticons are very useful in sentiment analysis. A recent gender based study [14] discussed that female users use more emoticons during the interaction in twitter as compared to males. Females also use first person singular more oftenly as compared to males in female to female interaction.

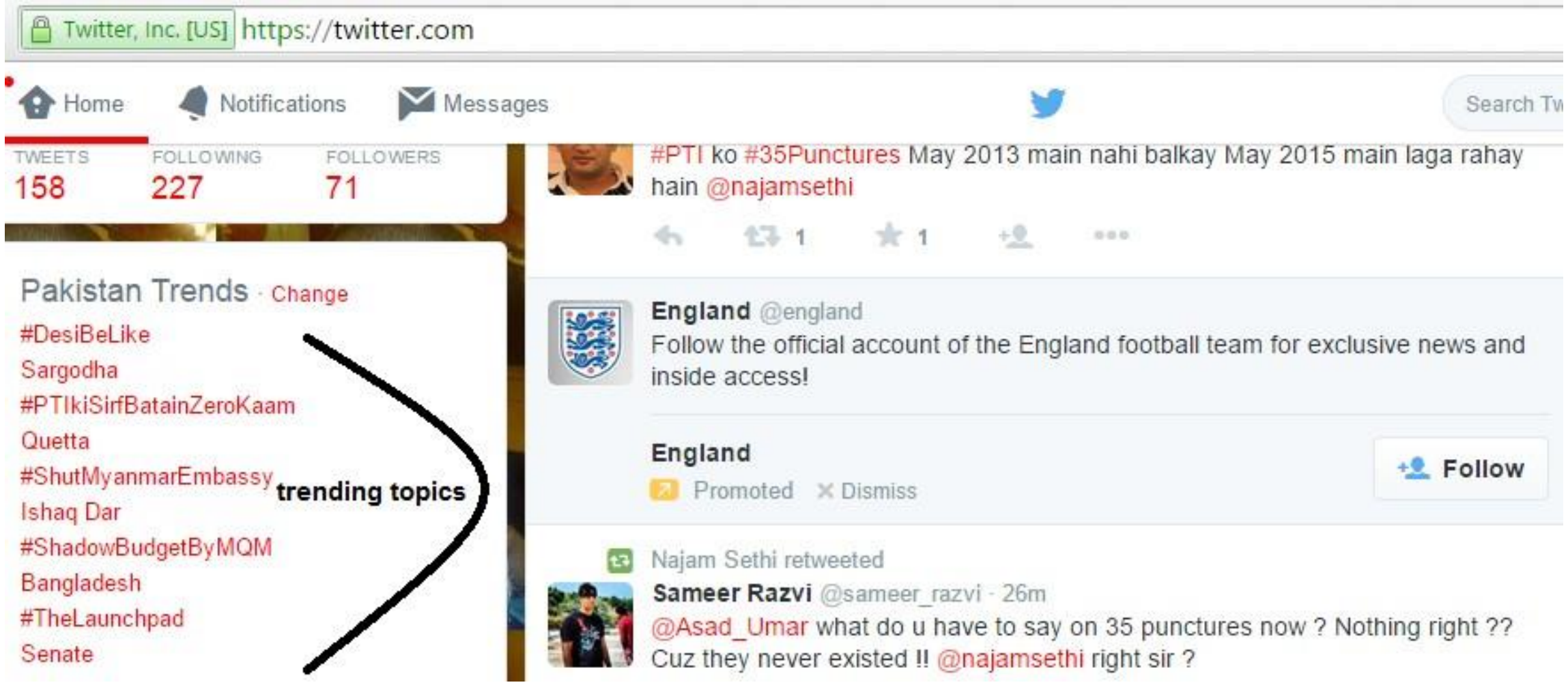

Figure 2.1: Trending topics on user's homepage

## 2.3 Twitter Terminologies

### 2.3.1 Follower

A follower is a twitter user, who subscribes to a user, so he may be able to receive updates from that user.

### 2.3.2 @Mention

To tag someone in a tweet, "@" symbol followed by user name is used. Figure 2.2 shows an example of mention.

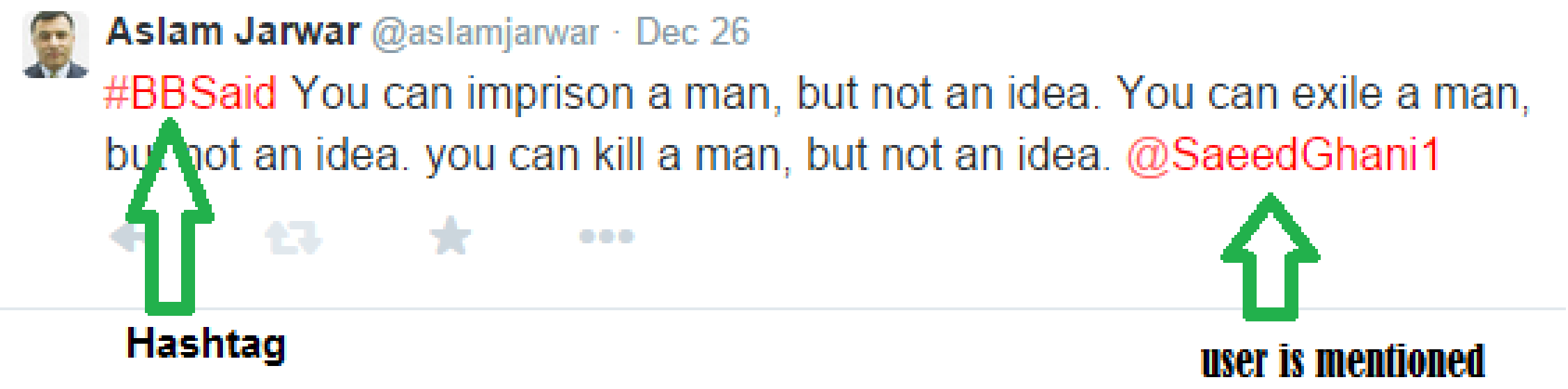

Figure 2.2: Example of mentioning of user-id in tweet

### 2.3.3 List

Twitter users create a single list or many lists of twitter users mostly related to each other in some aspect. By using lists, users can see a tailored stream of tweets of the users present in the list. The list can be private or public. The public lists related to person are shown in figure 2.3.

### 2.3.4 Hashtag (#)

A hashtag is a keyword or phrase that is prefixed with # sign, as with #azadimarch or #inqlabmarch. A Hashtag is used to categorize tweets according to its context, and it also helps users to search tweets of a specific topic.

### 2.3.5 DM

DM is a short of "Direct message", the message which is sent privately. Only the sender and receiver can see the message. A direct message gives the facility to users like private instant messaging. Twitter also supports sending direct message to a group of up to 50 people.

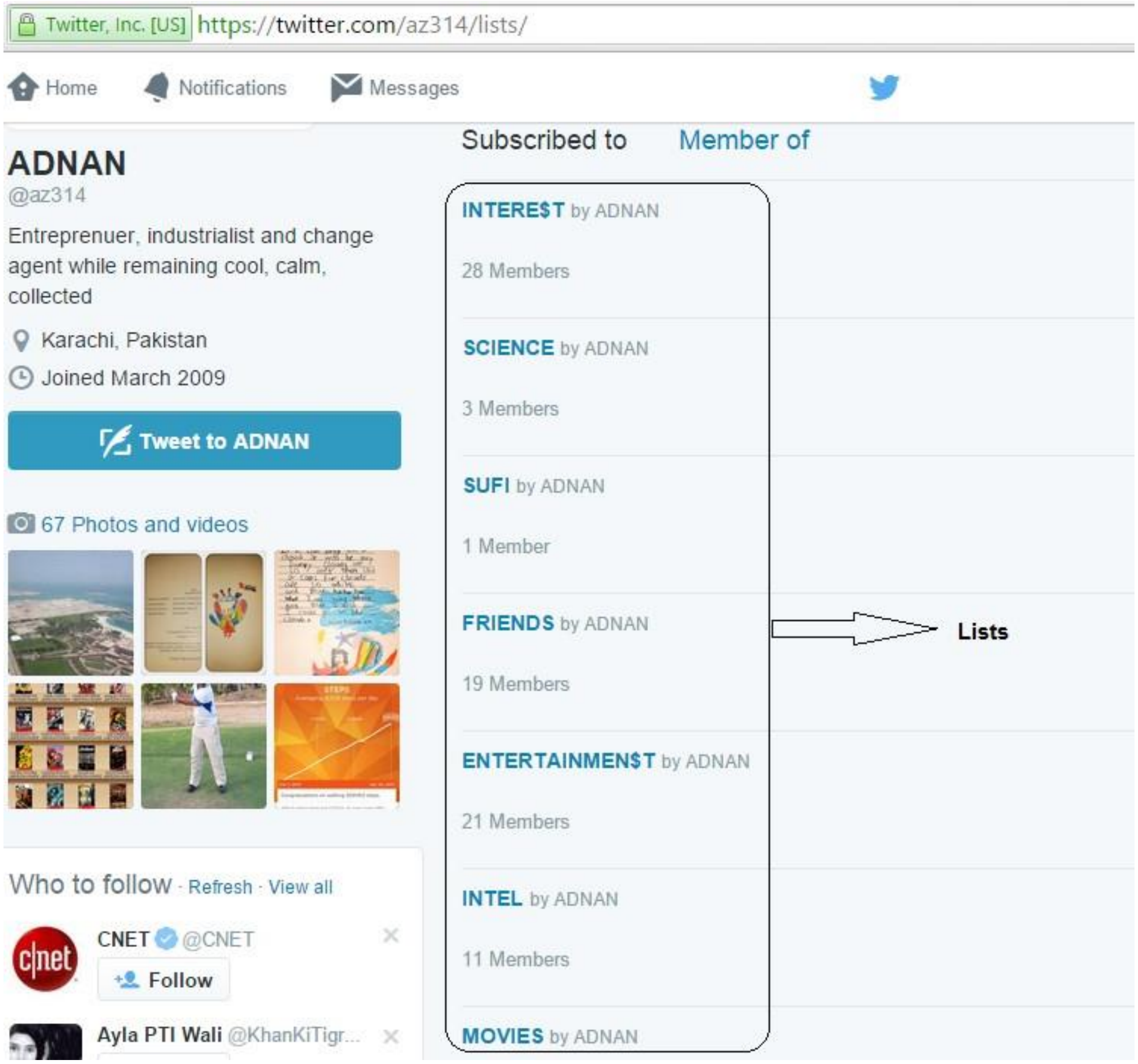

Figure 2.3: Public lists of a user

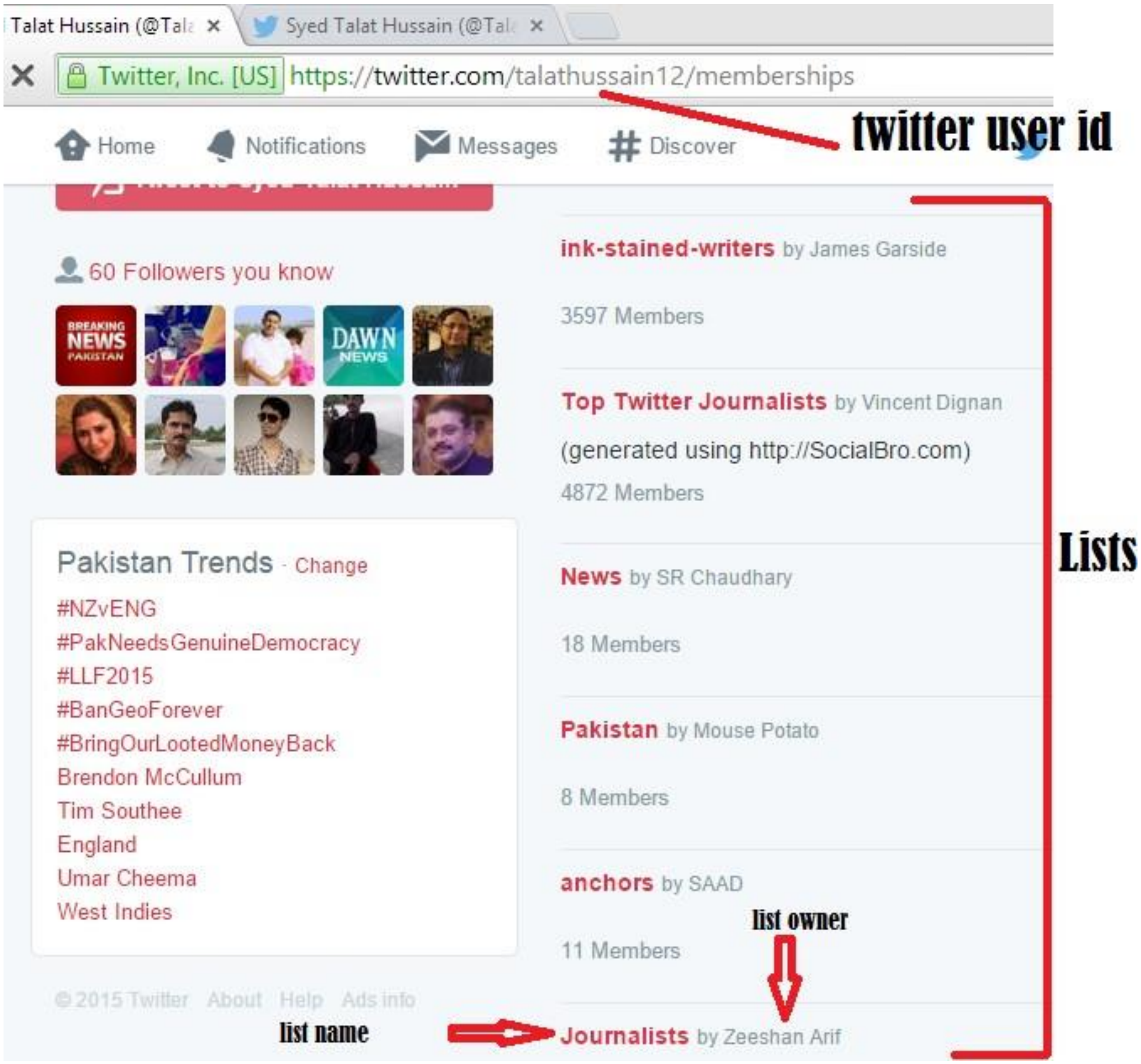

Figure 2.4: This figure shows lists which contain the twitter user id “talathussain12” of Senior Journalist Mr. Talat Hussain

### 2.3.6 URL Shortener

Due to limit of 140 characters people use short form of URLs in their tweets like bit.ly/1KMh16H. The one of popular online tool from Google is (https://goo.gl/)

### 2.3.7 Alert

Alert is type of message having a unique look in twitter and indicated with orange bell[2]. It facilitates public safety agencies to inform people during the emergencies. Figure 2.5 shows two alert messages.

### 2.3.8 Favorite

When a user likes a specific tweet of his or her choice by clicking on the star symbol below the tweet; is called a favorite as shown in figure 2.5.

### 2.3.9 Singleton

A tweet which is not a re-tweeted is called a singleton.

### 2.3.10 Pinned Tweets

Pinned tweets are displayed on top of a users profile page. Users pin tweets, due to the importance of their choice as shown in figure 2.6)

### 2.3.11 Private accounts

Twitter accounts by default are public. When a user configures its account as private, then the tweets from this account are not visible publically.

[2] https://media.twitter.com/best-practice/twitter-alerts

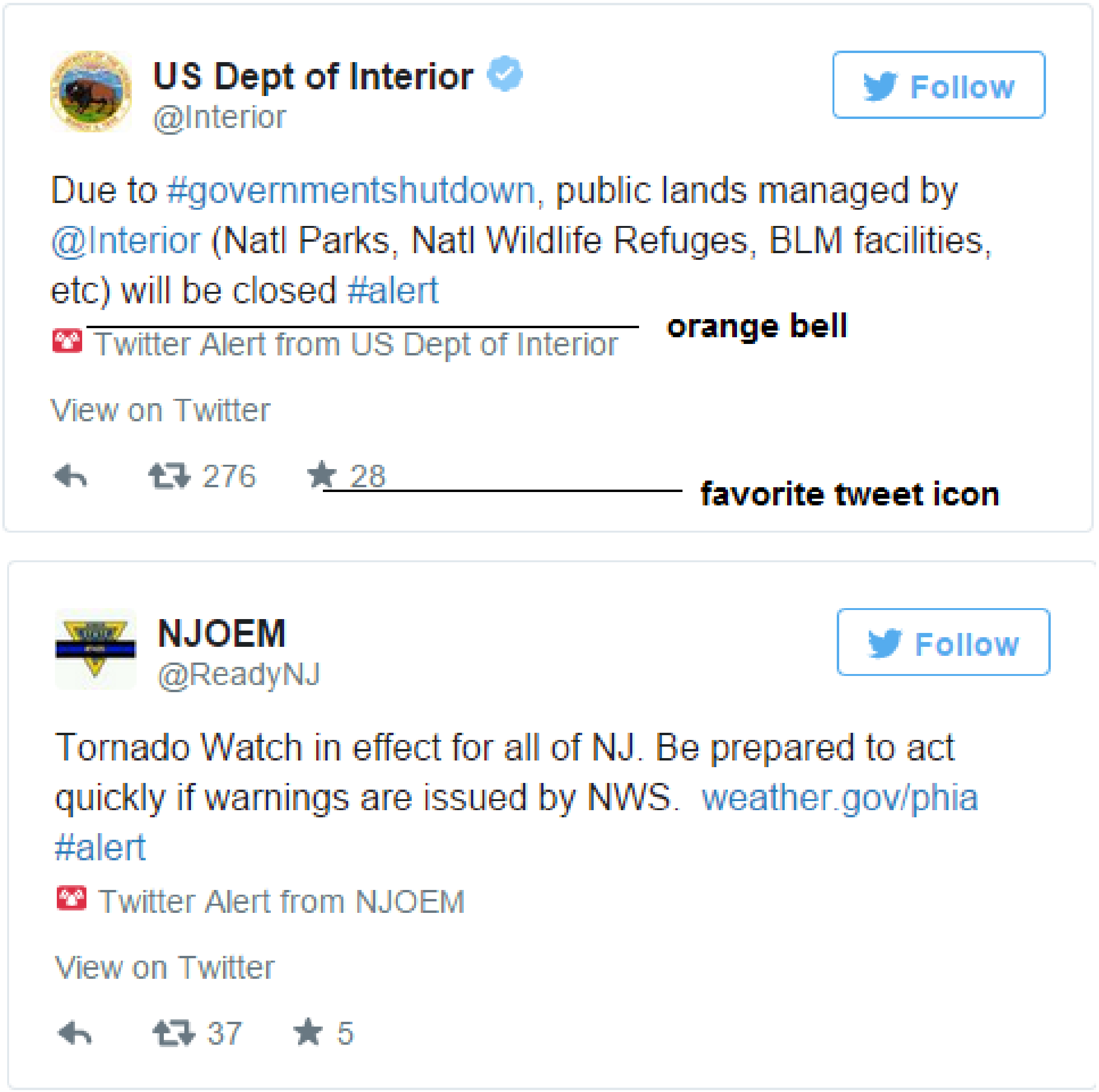

Figure 2.5: The figure shows alert messages indicated with orange bell and favorite indicated by star [a]

[a] https://media.twitter.com/best-practice/twitter-alerts

images/pinnedtweet.png

Figure 2.6: A tweet is pinned by user due to its high importance to him

### 2.3.12 Timeline

Time-line is a space on user's home page, where real-time stream of tweets are displayed in chronicle order.

## 2.4 Opinion

An opinion is point of view, expression and judgment about an entity consisting on facts or without facts, the entity can be any topic, event, or object[3]. Opinion is described in various literatures, which describes it in their own context. Computer science researchers consider the opinion as the sentence “subjectivity” [15]. According to a study [16], an opinion consists of four elements (opinion holder, opinion about the topic, sentiments about the topic and claim of opinion). For example in a sentence “I like black shirt”, I is the opinion holder, black shirt is topic, like is the positive opinion about the topic and the complete sentence is the claim.

## 2.5 Sentiments

Sentiment is defined as “an attitude, thought, or judgment prompted by feeling”[4]. Oxford dictionary defines sentiment as “A view or opinion that is held or expressed” and in another English language dictionary[5] “The thought or feeling intended to be conveyed by words, acts, gestures as distinguished from the words, acts, or gestures themselves”. [16] describes sentiment as the emotional part of expressed opinion. Sentiments can be positive, negative or neutral. When a person writes a letter, a blog or a micro-post, the selection of words and syntactic style of writing shows the peson's sentiment.

[3] http://www.abanet.org/buslaw/tribar/materials/20050120000000.pdf

[4] http://www.csc.ncsu.edu/faculty/healey/tweet_viz/, Last Accessed on 2015-02-23 11:38

[5] http://dictionary.reference.com/browse/sentiment

## 2.6 Emotions

Emotions are our subjective feelings and thoughts. Emotion studies are very broad in many fields, such as psychology, physiology, and sociology. Emotions can be expressed in many ways like facial expressions, gestures, different types of subjective experiences of an individual's state of mind. The scientists have categorized different types of emotions, these are love, joy, surprise, anger, sadness, and fear [17]. In psychology emotion is a phenomenon of complicated feelings, which influences the human behavior [18]. In natural language processing emotion classification from text is a growing field. In this field Paul Ekman's six basic classes of emotions (happiness, anger, disgust, fear, sadness, surprise) [19] are most popular and remain the focus of researchers. The emotions are also important to provide IoT based recommended services [20].

## 2.7 Are Opinions, Sentiments, Emotions the same?

In psychology and sociology; opinion, sentiment, and emotions are separate terms. But computer science researchers treat these three terms as same because these are the different stages of the same condition. Most of the studies deal "Opinion Mining" and "sentiment analysis" as the same phenomena and use them as synonyms [21]. In an article [22] use the terms opinion and sentiment interchangeably. When people give their opinion through written text, facial expression and spoken words, then their opinion is categorized into sentiments i.e negative, neutral or positive. After that sentiments are finer-granulated into positive emotions (happiness, likeness etc.) and negative emotions (anger, disgust, fear, sadness, mourn etc.). A recent study [23], addresses the issues of document level "sentiment analysis" and detect emotions in the categories of *"anger", "disgust", "fear", "happiness", "like", "sadness", "surprise" and "none"*. Therefore in our study, we use opinions, sentiments and emotions interchangeably.

## 2.8 Sentiment Analysis

The phrase of “sentiment analysis” first used in [24]. In textual natural language processing “Sentiment analysis” is the process of finding the opinion of the writer. “Sentiment analysis” is also known as “opinion mining” and it is a branch of “natural language processing” to study user’s opinions, attitudes and attributes [17]. [25] describes “sentiment analysis” or “opinion mining” as the study of knowing the people's opinions, attitudes and emotions towards objects. The object can represent a product, a topic or an event. There are many different levels of analysis for detecting the sentiments. These levels are:

### 2.8.1 Analysis at Document Level

Early research in sentiment analysis focused on document level analysis. Turney [26] and Pang [27] discussed and applied various methodologies to find the sentiment of the entire document. Document level analysis was helpful in the analysis of the product reviews and was also helpful in knowing whether people like or dislike a movie. It classifies the entire document into positive or negative opinion. In document level analysis; it is assumed that, the entire document is about a single entity. In research, it is observed that a single document about a single product may contain both positive and negative opinions, because the document contains many sentences. As an example, this paragraph: “I bought a phone two weeks ago. Everything was good initially. The voice was clear and the battery life was long, although it was a bit bulky. Then, it stopped working yesterday.” contains both positive and negative opinions.

### 2.8.2 Analysis at Sentence level

In this level of analysis; the sentence is classified as positive, negative, or neutral. The first stage is to identify that the sentence is subjective (sentence with opinion and views either positive or negative) or objective (neutral sentence) [28]. [29] finds the sentiments of tweets by classifying the tweets into subjective and objective. In this study authors observed that objective texts contains more common and proper nouns and subjective texts use personal pronouns more oftenly. The word's subjectivity depends in the context in which these words are used. Sometimes objective sentence contains the subjective opinion e.g. “We bought the car last month and the windshield wiper has fallen off.” The sentence level sentiment analysis is more challenging, because in a single sentence; mostly the context of the sentence is not described.

### 2.8.3 Analysis at object and Aspect level

The analysis types discussed in section 2.8.1 and 2.8.2 do not detect actual liking and disliking of a person from their textual information, because these levels miss the target of the opinion. Object or aspect level analysis is also called feature level summarization [30]. Aspect level analysis methodologies focus on the target of opinion and sentiment, without keeping concentration on sentence, phrase, paragraph and clause. In this analysis, the target of opinion helps in understanding the complexities of sentiment analysis. For example the tweet “core committee meeting of PTI starts, shah mehmood q, j tareen, s mazari, azam sawati... revolutionary of naya pakistan. Good joke” looks very positive, but the sentiments of the tweet is negative (i.e. criticizing tone). This sentence comprises three parts; the first part “core committee meeting of PTI starts” has neutral opinion, with target of opinion “meeting of PTI”. The second part “shah mehmood q, j tareen, s mazari, azam sawati... revolutionary of naya (new) pakistan” has positive aspect (i.e. revolutionary of naya (new) pakistan). The third and last part of the tweet “Good Joke” has positive sentiment, but if we analyze this part with the target of opinion, the overall sentiment of the tweet is negative. Aspect level analysis also shows the summary of opinions about objects in the sentence and gives the overall aspect of the sentence.

## 2.9 Applications of Sentiment Analysis

Opinions and sentiments have a very effective role in completion of our routine and new tasks in everyday of life. When we face some strange situations, we get opinion from our dear and near ones. Parents always take care of the sentiments of their children, because the sentiments influence our behavior. Political parties experts mine public opinion from social media and detect sentiments and use these sentiments for their political activism. Sentiments may also influence the election and used in predicting election results [31,32]. Researchers also propose and develop many applications for sentiment analysis. [33] develop a system for Irish election 2011, the system monitors political sentiments from live twitter data and it predicts the election results. Yang Liu and others propose a sentiment analysis model for sales indicators, this model mines the data from blogs where people share their views about the product. They conducted their experiments on a movie data-set [34]. Johan Bollen and others used twitter data to predict stock market [35]. Some online sentiment analysis applications are also available. "Sentiment140"[6] extracts sentiments from twitter data about a product. In figure 2.7 sentiments of "Nokia" are shown percentage wise. Sentiment analysis is also used in the application "Weather Sentiment Prediction"[7] to predict weather sentiments. This application mines data from social media against a specific area and classifies the data into positive and negative sentiments as shown in figure 2.8.

[6]http://www.sentiment140.com/ Last accessed on 2015-03-01 23:43

[7]http://www.sproutloop.com/prediction_demo/ Last accessed on 2015-06-08 23:43

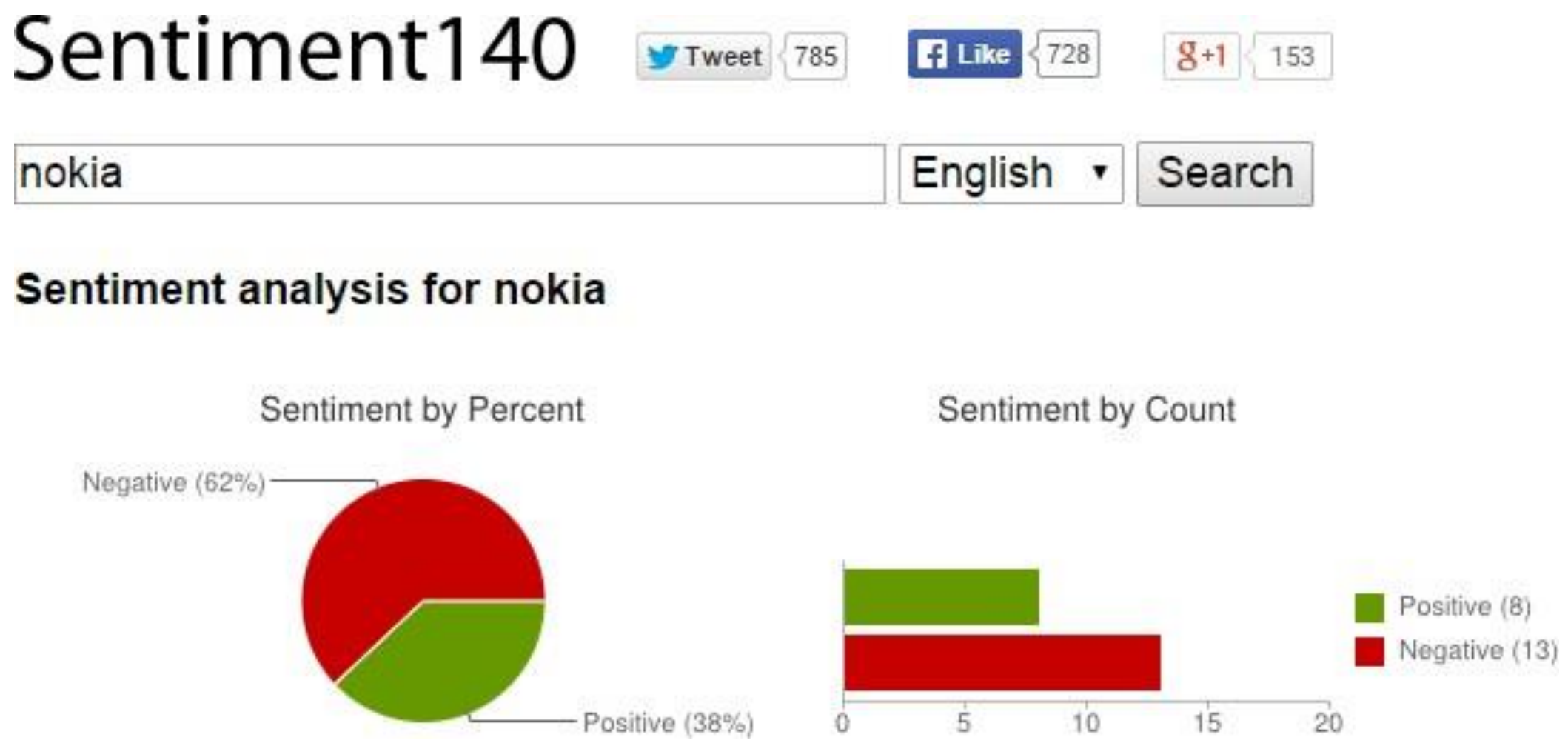

Figure 2.7: “Sentiment140” displaying the sentiments of Nokia in percentage.
[a]

[a]http://www.sentiment140.com/search?query=nokia&hl=en

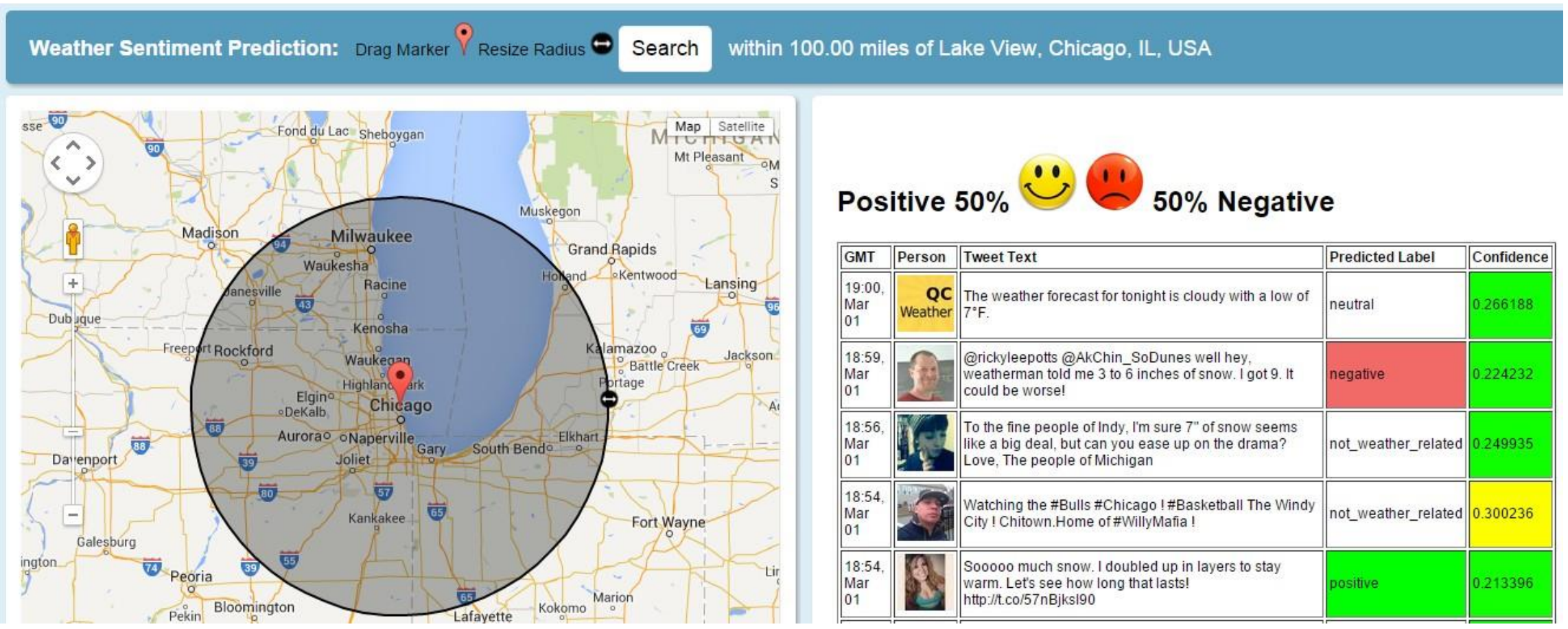

Figure 2.8: Shows weather sentiments of within 100.00 miles of Lake View, Chicago, IL, USA.
[a]

[a]http://www.sproutloop.com/prediction_demo/

### 2.9.1 Existing Solutions for Sentiment Analysis

In sentiment analysis, the first step is the selection of text features. These features include the frequency of terms present, use of adjectives (parts of speech), phrases, and negations for example not sweet, not intelligent, not harsh [36]. The features are selected using two methods; lexicon-based and on the basis of statistical measures. The lexicon-based method is a manual method, in which the human annotator annotates the features manually. The statistical method is fully automatic and widely used [25]. However the chances of novice features are much greater as compared with lexical method.

The steps involved in features selection are:

- **Sarcasm:** Identifying the contextual meaning of the verb used in text. For example *"Well, this day was a total waste of makeup."*

- **Anaphora:** In the text users use noun and pronoun in a way, which causes difficulties to know the real sense of these. For example in sentence *"We had a lavish dinner and went for a walk, it was awful."* the meaning of "it" is ambiguous; whether "it" refer to "lavish dinner" or "went for a walk".

- **Abbreviations:** In text and mostly in tweets, abbreviations are used frequently, so for feature extraction and sentiment analysis the complete word is required. Researchers have used dictionaries; for the resolution of abbreviations [37].

- **Stemming:** In feature selection, it is necessary to extract the root word. For example cars to car, flies to fly and etc.

Machine learning, and lexicon based sentiment analysis are two main methodologies [25] used by researchers for detecting the sentiments from extracted features. In machine learning suitable algorithms based on linguistic features are used. Lexicon based approach as use precompiled sentiments of terms. [38], identifies four feature selection methods (Mutual information, Information Gain, Chi-square and Document Frequency) and five learning methods (centroid classifier, K-nearest neighbor, winnow classifier, Naive Bayes and Support vector machines) and observe that best classifier depends on domain topics. There are also hybrid approaches [37], which use both methodologies for maximizing the accuracy of the analysis. In machine learning methodology, supervised and unsupervised methods are used. In supervised method a large number of labeled trained documents is used, whereas in unsupervised method, documents with no labels are used [25]. Unsupervised methods overcome the difficulties of labeling the trained documents.

Stanford CoreNLP [39] and Synesketch [40] are open source sentiment analysis libraries. Stanford CoreNLP is a java based toolkit, which is helpful in natural language processing and analysis. Some important CoreNLP supported tools are discussed in table 2.1. In CoreNLP sentiment annotator uses the supervised methodology to train the classifier. For more accurate results the classifier uses the sentiment treebank, this includes 215,154 phrases labeled with fine-grained sentiment labels in 11,855 parse tree sentences. “Recursive Neural Tensor Network” is used to reduce the complexity in sentiment composition. This sentiment annotator classifies the sentence into positive/negative polarity with accuracy ranging from 80% to 85.4% and in fine-grained classification up to 80.7% [43]. In fine-grained classification, the sentence is classified in 0 to 4 integer values (i.e. “Very Negative”, “Negative”, “Neutral”, “Positive” and “Very Positive”). Figure 2.9 shows the “Recursive Neural Tensor Network” of a tweet with “very negative” polarity.

Synesketch uses the word lexicon based on WordNet, lexicon of emoticons, common slang words and a set of heuristic rules for extracting the fine-grained Ekman emotion classes (i.e. happiness, sadness, anger, fear, disgust, and surprise) and classifies the sentence in positive with value (+1) and negative with value (-1) polarity [40].

| Tools | Description | Example |
|---|---|---|
| Tokenize | breaks the text into tokens | sentence “Karachi is the biggest city of Pakistan and it is located on the Arabian Sea coastline” broke into sixteen token |
| sspilt | breaks the sequence of tokens into sentences | |
| pos | tags the tokens of sentence to parts of speech (POS) | Tokens from above sentence tagged as Karachi=NNP, is= VBZ, the=DT, biggest=JJS, city=NN, of=IN, Pak-istan=NNP, and=CC, it =PRP, is=VBZ, located=JJ, on=IN, the=DT, Arabian=NNP, sea=NNP and costline=NN for tags see Penn Treebank Project [41] |
| lemma | generates the lemma of words | The lemma generates the above sentence as “Karachi be the big city of Pakistan and it be locate on the Arabian Sea coastline” |
| ner | identifies names of locations, organizations and na- tionalities from text | In the above sentence Karachi, Pakistan, Arabian and sea will be tagged as location |
| parse | provides complete support of syntactic analysis with both constituent and dependency representation, based on a “probabilistic parser” [42] | |
| sentime nt | detects sentiments from text | The above sentence has negative sentiment |

Table 2.1: CoreNLP Natural Language Processing supported tools

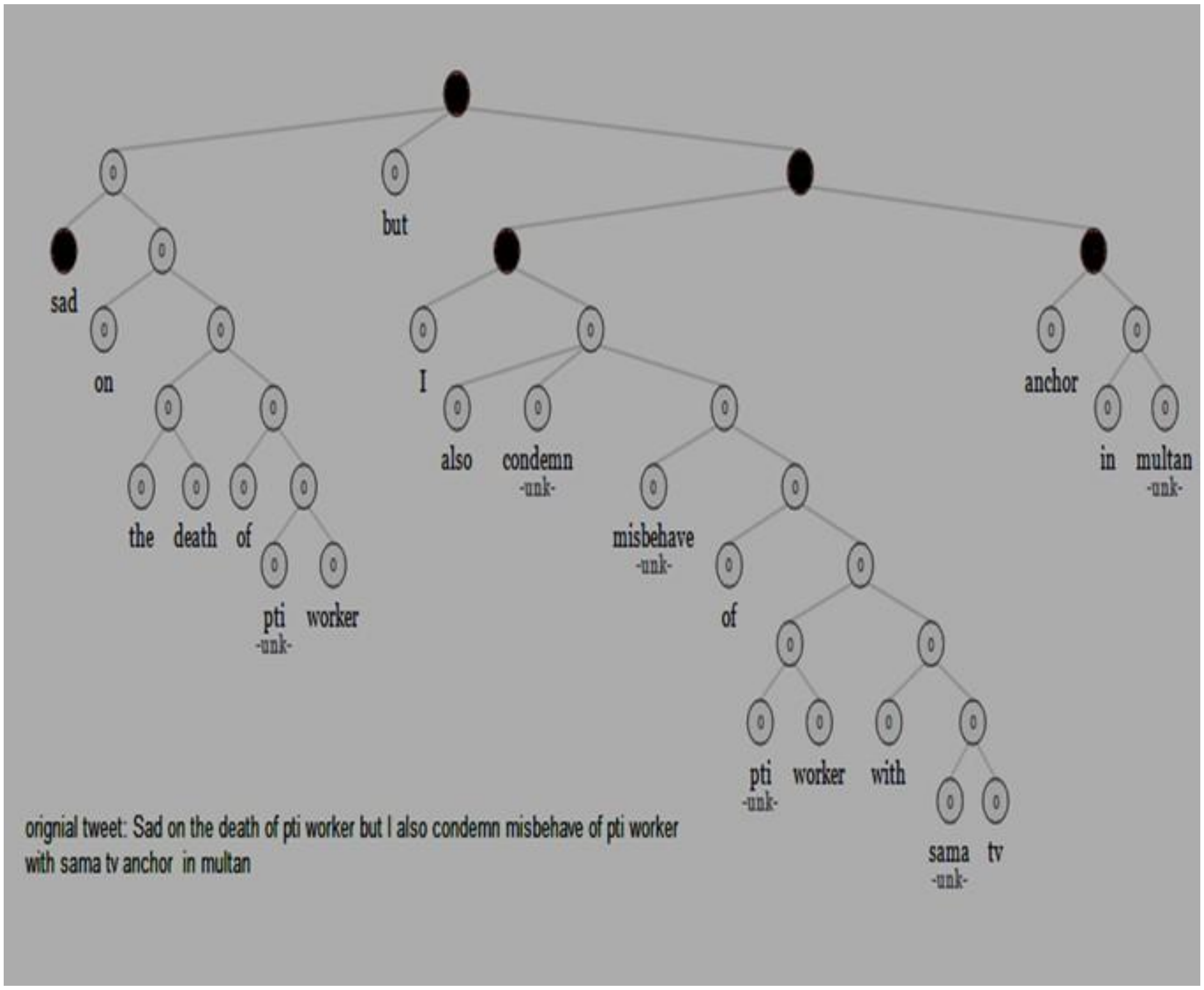

Figure 2.9: Sentiment of a tweet in “Recursive Neural Tensor Network”. Here black holes mean “negative” and grey holes mean neutral.

## 2.10 Snowball Sampling

Snowball sampling or "chain-referral sampling is a non-probabilistic sampling technique where existing study subjects recruit future subjects from among their acquaintances". This technique is used for identifying specific communities in twitter like celebrities, media, organizations, and blogs [44]. [44] starts with seed users $u_0$ belonging to a specific community. The seed users $u_0$ are mostly famous personalities within a community. All lists having the users $u_0$ are retrieved. The retrieved lists for each community are filtered on the basis of manually chosen keywords. The filtered lists $l_0$ contain only those lists, whose names match, contained manually chosen keywords. The process of snowball sampling is shown in figure 2.10. In a recent

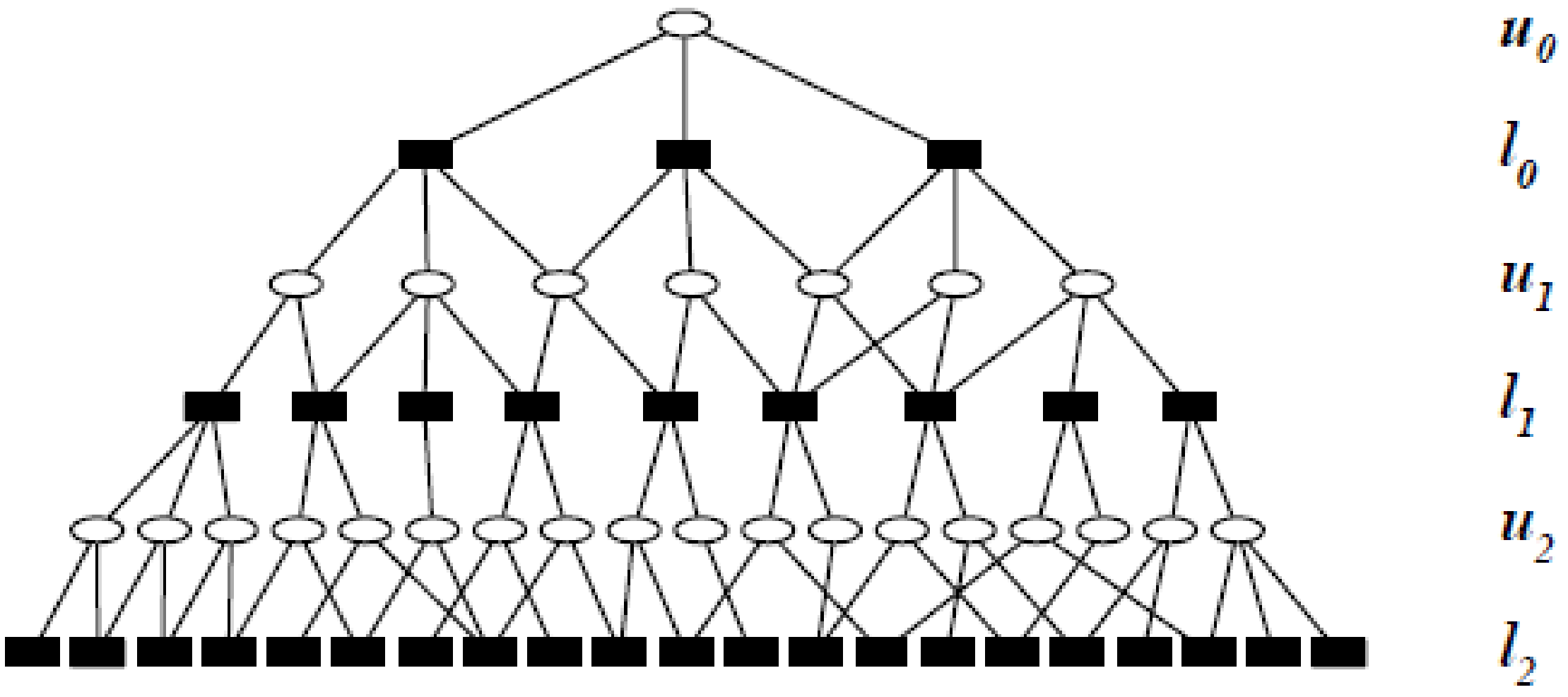

Figure 2.10: The diagram shows the process of snowball sampling, which is used for identifying the members of a specific community. In the diagram, circles represent users (e.g., $u_0$, $u_1$, $u_2$) and squares represent lists (e.g., $l_0$, $l_1$, $l_2$). Initially seed users $u_0$ are searched in the lists and associated set of lists $l_0$ is retrieved. In the second stage $l_0$, is used for getting new users $u_1$. Further new lists $l_1$ are retrieved from users $u_1$. This process continues until the required number of members of a community is crawled. [44]

study [37], the researchers have used snowball sampling for spell checking in English tweets.

## 2.11 Evaluation Methodologies

Evaluation is a critical part of any research. Two types of evaluation methods, i.e. quantitative and qualitative are widely used. In the field of opinion mining and sentiment analysis; precision, recall and f-measure are the most popular evolutionary techniques. Many recent studies [37] [45,46] used precision, recall and f-measure techniques for evaluation. In emotion watch prototype, [47] adopts user study evaluation technique and also performed in-depth, qualitative, and formative evaluation.

## 2.12 Summary

In this chapter we discussed the literature related to our study, and overall to sentiment analysis. We define microblogging, overview of twitter, definitions of opinions, emotions, and sentiments and further discussed that opinions, emotions, and sentiments are different names to same entity at least from computational aspect. We also discussed snowball sampling and evaluation techniques related to our study. In next chapter we will discuss our proposed framework.

# Chapter 3

# Proposed Framework

This chapter discusses the proposed framework of our study. In section 3.1, we give the introduction of our proposed framework. Section 3.1.1, discusses the first part of our framework for identifying persons of a particular community in twitter. Section 3.1.2, presents the second part of framework, in which tweets of the targeted community are retrieved and filtered. Section 3.1.3 discusses the third and last part of our framework, which is about tweets cleaning and sentiment detection.

## 3.1 Proposed Framework

As discussed in sections 1.1 and 1.2, there are number of people having different life styles, professional fields and communities. They express their opinions in social media on different events occurring in their surroundings.

How can we know the sentiments of a particular community about an event. To answer this question we propose a generic framework which can be used for knowing the sentiments of any specific community.

Our proposed framework consists of three parts. Each part has its sub parts. Figure 3.1 shows our proposed framework. We briefly discuss each part in following sections.

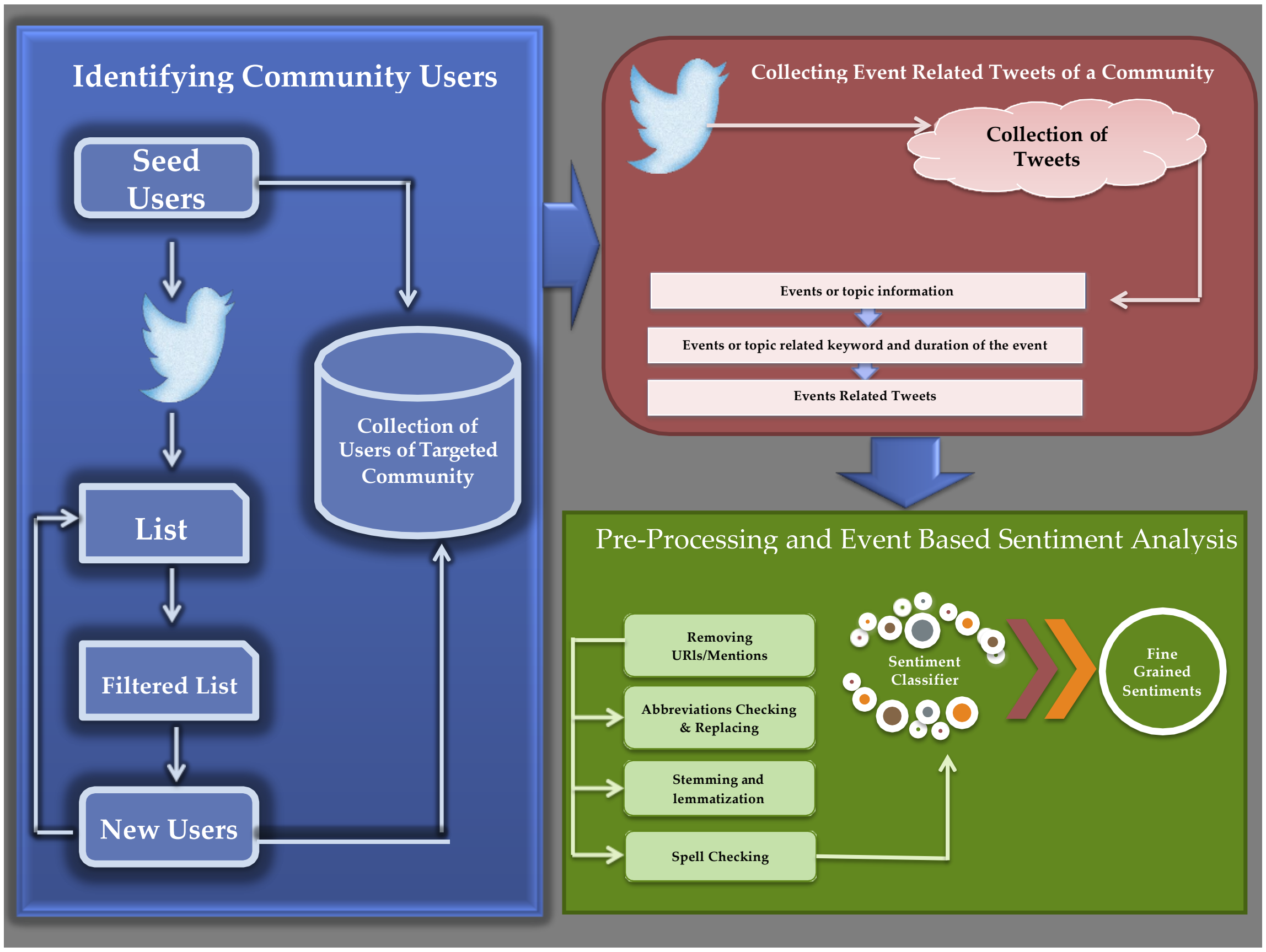

Figure 3.1: Detailed Architecture of Proposed Framework

### 3.1.1 Identifying Community Users

This is the first and starting point of our proposed framework. This part uses the twitter Rest API and snowball sampling technique (as discussed in section 2.10), for retrieving twitter user's ids of targeted community.

We manually identify seed users, of a community. The seed users are well known representatives of a community. We retrieve these all the lists (as discussed in section 2.3.3) containing at least one of the seed users. These lists contain other users of the same community. But they also contain users which do not belong to the community. To exclude such users, lists are filtered with keywords for removing irrelevant users. In such way, new relevant users are retrieved from these filtered lists and are used to get new lists. This process is repeated until a sufficient number of users are retrieved.

In the final stage of this part, users in all the filtered lists, including the seed users are collected in a single repository. These users represent a community and are used in the next part of the framework.

### 3.1.2 Collecting event related tweets of a community

We require a data-set for fine-grained sentiments analysis of the tweets pertaining to the targeted community. The second part of the proposed framework acquires the tweets of the targeted community and filter these tweets based on events.

In this part of the framework, first tweets of all the users of the targeted community are retrieved, using twitter API. While downloading the tweets, API limits imposed by twitter should be respected. If rate limits are not respected then twitter blocks the IP address for indeterminate time. Our framework automatically adjusts twitter time and rate limit for easy downloading of tweets.

To know the sentiments of the targeted community about any specified event or situation occurred in their surroundings, it is necessary that we filter tweets in our data-set with respect to the events of interest. For filtering event related tweets from data-set we use seed words related to the event and the duration of the event.

The seed words might not be sufficient to retrieve most of the tweets belonging to an event. To fetch most of the tweets related to an event, we first retrieve all the tweets containing the seed words. Afterwards we identify most frequent words in these tweets and use these frequent words along with the seed words to fetch most of the tweets related to the event.

### 3.1.3 Event based Sentiment Analysis

This is the last and third part of our proposed framework, the pre-processing, cleaning, sentiment detection and fine-grained classification related tasks are performed. As in our literature review, we have discussed that the tweets contain un-structured text and also contain metadata. But the standard sentiment analysis tools do not understand these conventions. If we do not pre-process the tweets, accuracy of the sentiment analysis tool may be effected negatively.

Therefore pre-processing is an important task which involves:

- Removing of URLs
- Removing of mentioned user ids
- Slang words and abbreviations detection and replacing them with appropriate words
- Words stemming and spell checking

Sentiments of all the tweets related to a specific event are analyzed using a sentiment analysis tool. The sentiment analysis tool takes the refined tweets as input, detects sentiments and assigns the tweets fine-grained sentiment classes.

The results of the sentiment analysis reflect the sentiments of a community about a particular event. We can also use these results for in-depth analysis, like which community members having positive, negative, or neutral sentiments about an event.

## 3.2 Summary

In this chapter we proposed a framework for sentiment analysis of a community. Different parts of the framework were discussed in detail. In next chapter we discuss the experiments, dataset and evaluation of the proposed framework.

# Chapter 4

# Experiment and Evaluation

In this chapter we discuss the dataset, experimental setup and the results to evaluate the proposed framework. Section 4.1, gives the introduction of tools and libraries used in our experiments. In sections 4.3, 4.4, and 4.5 we brief the experiments for testing the proposed framework. Section 4.6, gives overview of the dataset. Section 4.7 presents the results of our experiment. The analysis of sentiments is discusses in section 4.8. The section 4.9 highlights event wise sentiment classification and the overall discussion of the chapter is given in section 4.10.

## 4.1 Tools-APIs-Libraries

We use following APIs and libraries in our study.

1. LINQ to Twitter[1]
2. Twitter4J[2]

**LINQ to Twitter:-** It is an open source library written in C#. This library is useful for connecting .NET applications with the Twitter REST API. LINQ to Twitter uses the LINQ syntax for queries. JSON[3] formatting and parsing is also supported by this

[1]http://linqtotwitter.codeplex.com/
[2]http://twitter4j.org/en/index.h
tml [3]http://json.org/

API. OAuth[4] mechanism is used for connecting with twitter. In our experiment, we use this library for collecting twitter lists, list owners, members, and tweets' language and location.

**Twitter4J:-** It is an open source library implemented in Java. It supports OAuth and JSON data parsing. This library is widely used in Java for using twitter API, without any additional dependency. Due to limitations of data synchronization and retrieval, we use twitter4J in our framework instead of LINQ to twitter for retrieving tweets of a targeted community.

## 4.2 Case Study

The journalist community actively uses social media services like twitter for expressing its opinion and point of view regarding the events and news occurring in their surroundings. Many people follow journalist for the latest information and are interested to know their opinions on the issues or events happening around them. The journalists mostly keep their twitter profile and tweets publically accessible to all the users. The journalist community plays an influential role in governmental policies and public affairs, because information flows from them to common citizens.

Therefore we choose journalists as our targeted community and our dataset consists of the tweets of Pakistani journalists and we will use this dataset for testing our hypotheses and proposed framework.

To know the opinion of journalist community on any topic or event, we need user ids of journalists from every part of the country, from Karachi to Kashmir, including the remote areas of Baluchistan and FATA. Once we have information about journalist, we can analyze their tweets for further analysis.

[4] http://oauth.net/

## 4.3 Retrieving Twitter ids of Journalists

We started with three twitter ids of renowned Pakistani journalists (HassanNisar[5], HamidMirGEO[6] and TalatHussain12[7]) as seed users using snowball sampling technique for retrieving users represent the journalist community. In first phase by referencing these seed users, we retrieved 7477 lists created by different users that contain at least one of the seed users. These 7477 lists contained 2257 unique labels. We manually filtered relevant lists, and were left with 166 unique labels shown in table 4.1. Labels in table 4.1 are in English, Urdu and Roman Urdu like "Sahafi" which means journalist. Many labels in English have spelling mistakes, because English is not the national language of Pakistan.

The filtering was performed in the context of targeted community (in our case journalists) and on the basis of word frequency count. For example we retrieved lists having names *" -£l9î ęzIzlnj ,l,¥ , ,eþrg , .tlnZtl, Pak, Pak Politics, Pakistan, Pakistan News, Pakistani Politics, PK-Politics, the-journalists, anchor, IMPORTANT, Inqalabi Group, Journalist, Anchors, Journos, Analysts, Pakistan, Pakistani, Journalists Private List, TOP Columnist"* and out of these example lists, only the lists having ,eþrg, Journalists and Anchors in their labels were selected for next step.

As explained above that 7477 lists belonged to different users. After using 166 unique labels, there were 592 lists of different users including duplicate labels. Now by using the noise free and filtered 592 lists of different owners, we obtained 15390 new members of the targeted community. Further we filtered semi manually the data of 15390 new members by targeting Pakistani journalists on the basis of their locations (as mentioned in twitter users profiles) and profile description. After filtering we obtained 923 users of Pakistani journalists community. The remaining 14467 users are not relevant to Pakistani journalist and mainly contain users belonging to different news agencies and electronic media. The process may be continued to increase the size of the focused community.

[5] https://twitter.com/HassanNisar
[6] https://twitter.com/HamidMirGEO
[7] https://twitter.com/TalatHussain12

## 4.4 Collecting and filtering tweets into events

**Collecting Tweets:-**

Our proposed framework completed the first task of getting twitter ids of Pakistani journalists. Now we need to collect tweets of 923 Pakistani journalists. Twitter provides a single method for collecting last 3200[8] tweets of a user, with meta-data in JSON format. However twitter imposes restriction of time limit on downloading data and API calls. To handle time limits and saving time from parsing JSON data, we used an open source java library Twiter4J[9].

In order to get maximum API requests in less time, we used four twitter application keys in four different threads for tweets collections. We collected 2,107,374 tweets with meta-data (language, location, place and posting date) of 923 Pakistani journalists, in the duration of six days from January 18, 2015 to January 24, 2015.

**Choosing Events:-** To analyze the sentiments of journalist's community, we need to analyze their tweets for specific events. We used Wikipedia[10] to get the list of events related to Pakistan. The list contains basic information (i.e. domain, location, start and end time, involved objects, targeted objects) of events. The list is useful for selecting keywords for searching and categorizing tweets based on specific events. Our list includes information of seven events (i.e. Jinnah International Airport attack[11], Operation Zarb-e-Azb[12], Azadi March/Inqlab March[13], NA-149 Multan By Election 2014 [14], 2014 Men's Hockey Champions Trophy[15], Peshawar school attack 2014[16], and Twenty-first Amendment to the Constitution of Pakistan/ miltary courts[17]) related to Pakistan .

[8]https://dev.twitter.com/rest/reference/get/statuses/user_timeline [9]http://twitter4j.org/en/index.html
[10]http://en.wikipedia.org/wiki/2014_in_Pakistan
[11]http://en.wikipedia.org/wiki/2014_Jinnah_International_Airport_attack [12]http://en.wikipedia.org/wiki/Operation_Zarb-e-Azb
[13]http://en.wikipedia.org/wiki/Azadi_March

Table 4.1: Twitter lists from which Pakistani Journalists twitter ids were retrieved

| **list names** | **list names** | **list names** | **list names** |
|---|---|---|---|
| achors | Jornalists | MediaPerson | Sahi-fav |
| analysist | Jounalism | Media-person | saifi |
| Analysts | Jounalists | media-personals | sr. journalist |
| analysts & journalists | jounerlisum | Media-persons/Journalists | the e-journalists |
| Analysts opinion makers | Journal lists | MyList-Journalists | TOP Columnist |
| Anch / host / journalist | Journalism | National Journalists | Top editors |
| anchars | JOURNALISM. | National-Journalist | TOP Journalist |
| Ancher | Journalist | News Analysts | Top journalists |
| Anchers | journalist circle | News Anchor | TopJournalists |
| Anchor | Journalist II | NEWS ANCHORE | tv analiasts |
| anchor person | Journalist/anchors | News Anchors | Tv Anchors |
| Anchor person PakTv | Journalist-and-Analyst | News-anchor | TV Anchors & Journalists |
| anchor persons | Journalist-n-Anchors | NewsAnchors | Tv ancors |
| Anchor Prs | Journalist-of-pak | News-anchors | TV Ankers |
| ANCHORES | JournalistPK | News-Ancor | TV Experts |
| AnchorJournalistColumnist | Journalists | NewsAndJournalism | TV HOST |
| anchor-person | Journa-Lists | News-journalist | TV, Anchor |
| Anchorpersons | Journalists & Anchors | Pak Journalist | Tv-Anchors |
| Anchor-Persons | Journalists & Editors | Pak Journalists | TV-Anchors-and-Jurnalists |
| Anchors | Journalists (Pakistan) | Pak Journo | [illegible] (TV anchors) |
| Anchors & Analysts | Journalists / Anchors | Pak Journos | [illegible](Journalists) |
| Anchors & Reporters | journalists and analysts | Pak Journos & Analysts | |
| Anchors and Journalists | Journalists of Pakistan | Pak News Anchor | |
| Anchors List | journalists/anchors | Pak News Anchors | |
| Anchors PK | Journalists-Anchers | Paki-Anchor-Persons | |
| Anchors.PK | Journalists-Domestic | Paki-Sahafi | |
| Anchors/Jornoulists | Journalist-Writers | Pakistan Journalism | |
| Anchors/Journalist | Journals & Anchors | Pakistan Journalists | |
| Anchors/Journalists | journalst | Pakistan Journlists | |
| Anchors-Journalists | journelist | Pakistan Journos | |
| Anchorts & Hosts | Journilists | Pakistan-Ancoer | |
| Ancor persons | Journlist | Pakistan-gernalist | |
| Ancors | journlists | Pakistani Ancher | |
| Ankar-person | Journo List | Pakistani Anchors | |
| anker | Journo watch | Pakistani Jounalists | |
| Anker Person | Journolist-PK | pakistani journalist | |
| ankers | Journolists | Pakistani Journalists | |
| Columist | Journslist | Pakistani Journos/Newsmen | |
| Columnist | juornlist | Pakistani Media Personel | |
| Columnist1 | jurnalis | Pakistani-journalist | |
| columnists | jurnalist | Pakistan-Journalist | |
| common list | Jurnlists | Pakistan-News-anchors | |
| Famous journalist | Local Journalists | PakJournalists | |
| Favorite-columnists | Media Parson | PakJournos | |
| Favourite Journalist | Media People | PAK-MEDIA PROFESSIONALS | |
| Favourite Journalists | Media Person | pk journalists | |
| Femala Journlist | media person,s | pkJournalists | |
| general list | Media Personalities | PK-Journalists | |
| Generalism | Media Personnel | Popular Journalists | |
| Generalist | Media Personnels | Sahafat | |
| Generalists | Media persons | Sahafi | |
| Good Journalists | Media-Journalists | Sahafi’s | |

To test our hypotheses (section 1.3), we choose three events: Zarb-e-Azab, Azadi and Inqlab March, and 2014 Men's Hockey Champions Trophy from the list of seven events. We choose these events due to their different contexts. Zarb-e-Azab belongs to army operation against terrorists, Azadi and Inqlab March relates to politics and Hockey Champions Trophy is a sports event.

**Finding Event Specific Tweets:-** After having a list of events, next step is to identify tweets related to these events. Events specific tweets help us in knowing the opinion and sentiments of a targeted community against a chosen event. According to [11] most of the events or topics remain active in twitter for one week or less, but this is not always true as we will observe in our experiments. We detect events related tweets using different methods. We discuss these methods in the following subsections.

**Event Zarb-e-Azab** Zarb-e-Azab is a joint military operation conducted by Pakistan armed forces against Tehrik-i-Taliban Pakistan (TTP) and others militant groups in North Waziristan area. This operation started on 15 June 2014, and was continued till we start to retrieve tweets on 18 January 2015.

We searched tweets related to "operation Zarbe-e-Azab" from data-set with key-words found in the Wikipedia article "Zarb-e-Azb, Zarb-i-Azb, ZarbiAzb, ZarbeAzab, Zarbe Azab, North Waziristan operation" and limited tweets posted during one week from the starting date of the event. We initially found 760 tweets and 211 users who expressed their opinions on the event. For extended search of events related tweets, we

[14] http://en.wikipedia.org/wiki/Constituency_NA-149#By_Election_2014
[15] http://en.wikipedia.org/wiki/2014_Men
27s_Hockey_Champions_Trophy%
[16] http://en.wikipedia.org/wiki/2014_Peshawar_school_massacre
[17] http://en.wikipedia.org/wiki/Twenty-
first_Amendment_to_the_Constitution_of_

Pakistan

searched dataset without keywords, but limiting the search with one week of tweets creation time, and parsed all searched tweets into bags of words. By sorting the words by their frequencies we found new keywords “ZarbEAzb” and “IDPs” which were absent in Wikipedia article. Again we searched entire data-set with previous and new keywords without limiting the time period and detected 11693 relevant tweets with 609 users who participated in the opinion on the event.

**Event Azadi and Inqlab March** Azadi (freedom) and Inqilab (revolution) march launched by two political parties Pakistan Tehrik-e-Insaf (PTI) and Pakistan Awami tehrik (PAT) from Lahore to Islamabad for their demands from the government. Azadi March was demonstrated from 14th August 2014 to 17th December 2014 and Inqilab March from on 14th August 2014 to 21st October 2014.

The “Azadi March” is also called “Tsunami March”. This march was organized by a political party Pakistan Tehreek-e-Insaaf (PTI) under the leadership of party chairman Imran khan against the ruling party Pakistan Muslim League Nawaz (PML (N)) and Prime minster Mian Muhammad Nawaz Sharif. Imran Khan claimed that PML (N) manipulated the results of general elections 2013. So he plans for the march from Lahore to Islamabad in the month of August with a throng of his supporters. Initially Imran Khan named his march as “Tsunami March”, but later on he changed the name to “Azadi March” relating it with the day of independence of Pakistan 14th August.

The “Inqilab March” was organized and lead by the chief of Pakistan Awami Tehreek (PAT) Dr. Tahir-ul-Qadri. He claims that the current political system has totally failed to solve the problems and in providing the basic necessities to common citizens. Another reason was the incident of model town Lahore[18]. In this incident eleven PAT supporters died and around 100 injured in a violent clash with police[19]. Dr. Qadri lead the march with thousands of his supporters with the agenda that to get the resignation from the prime minister of Pakistan Mian Muhammad Nawaz Sharif and to introduce reforms for the welfare of common people and to launch a merit based governance system in the county.

[18]https://en.wikipedia.org/wiki/2014_Lahore_clash

[19]http://www.dawn.com/news/1128333

We searched our data-set with keywords obtained from Wikipedia article; “Tsunami, march, Azadi, azadee, Freedom march,dharna, Inqilab march, PTI, PAT,Zaman Park, Qadri, tuq, Imran, Qadri, Azadi Square,azadisquare” and limited the starting and ending time period. We found 107,186 tweets and 783 participated users. After using word frequency as did for the event “operation Zarbe-e-Azab” we found new keywords “khan, AzadiMarchPTI, PTIAzadiMarch, ImranKhanPTI, , ꜝ ɧ,a> , ꜝ t¥ parliament, TahirulQadri, ImranKhan, revolution, resignation, resign, container, AzadiSquare, containers, rigging and AzadiMarch”. After applying new keywords we detect 147,131 tweets with 796 participated users, but unlike the “Zarbe-e-Azab” event, we restricted our search to 125 days.

**Event Hockey Champions Trophy for Men 2014** This sport event was held from 6 to 14 December 2014 in Bhubaneswar, India. In the semi-final Pakistan won from India by 4-3 and Germany from Australia by 3-2. The final was won by Germany by defeating Pakistan by 2-0.

Similar to the previous two events, we searched for event specific tweets with keywords “hockey, hocky, games, trophy, champion” and limited search to tweets posted between 6 and 14 Dec 2014. We found 1,431 tweets and 276 people who participated in the discussion. By using word frequency of words we got new keywords “india, team, Pak, match, play, semi, final, players”. We searched data-set again with old and new keywords and detected 13,222 tweets specific to the event. In this event total 597 journalists participated.

## 4.5 Pre-Processing and Fine-grained Sentiment Analysis

**Tweets Pre-processing:** As discussed in sections 1.4 and 2.2 that tweets have limited length of 140 characters and this limited length forces users to use many slang words or abbreviations and emoticons. Twitter users also use user ids (mentions), URLs, special character (i.e. sign of exclamation!, double or single quotation) and hash-tags. Sometimes users also make spelling mistakes. These twitter specific features may affect the accuracy of popular sentiment analysis tools. As popular sentiment analysis tools are not specially built for twitter.

Twitter specific features should be handled before extracting features for sentiment analysis. Addressing these features will facilitate us in achieving better accuracy [48]. Following operations are performed on all the tweets fetched in the previous step.

- **URLs:** URLS are removed by using the following regular expression:

```
\b(?:https?://|www\.)\S+\b
```

- **Mentions (user ids):** Twitter user ids are removed by using the following regular expression:

```
\b(?:\@?)\S+\b
```

- **Hashtags:** Hashtags are detected and # symbols are removed from words because without removing hash symbol sentiment analyzer could not check words polarity. The word following by the hash symbol is kept.

- **Slang words or abbreviations:** Slang words or abbreviations are detected by checking English words in WordNet[20] dictionary. If word not found in dic-tionary, it means that word is not of English language or the word is an abbreviation. Then the system checks these words from customized Netlingo[21] acronyms dictionary. The slang words are replaced in tweets, if it is found in the slang dictionary. The customized acronyms dictionary is shown in table 4.2.

[20]https://wordnet.princeton.edu/

- **Lemmatization:** Root words are detected and replaced in tweets by using lemasharp library as discussed in section 2.9.1.

Finally refined tweets are passed to sentiment analyzer for knowing the opinions expressed in the tweets.

**Detecting Sentiments-:** We have used Stanford CoreNLP sentiment annotator toolkit and Synesketch sentiment classifier engine as discussed in section 2.9.1. CoreNLP toolkit classifies a phrase into five integer values ranging from 0 to 4. 0 to 4 values describe the fine-grained sentiment classes i.e. "Very Negative" (0), "Negative" (1), "Neutral" (2), "Positive" (3), and "Very Positive" (4).

Each event related pre-processed tweet is passed to the Stanford CoreNLP sentiment toolkit and Synesketch engine for tagging the fine-grained sentiment class. Based on empirical analysis, we find CoreNLP toolkit more accurate than the Synesketch.

For example the tweet "RT @NazishMh: Nobody at peace tonight Pakistan up and cursing its leaders for landing our beloved country in this MESS #IK #TUQ" classified by synesketch as positive and CoreNLP annotated it as very negative. More examples of both classifiers are shown in table 4.3

## 4.6 Dataset

Our complete dataset includes 27,110 twitter lists belonging to 21,884 owners. These lists contain twitter ids of 16,915 journalists including at least, 923 Pakistani jour-nalists. This dataset contains 2,107,374 tweets with meta-data (i.e. tweet language, location and tweet posting time).

[21] http://www.netlingo.com/acronyms.php

Table 4.2: Customized Acronyms Dictionary

| S.No | Abbreviation/Slang | Full Form |
|---|---|---|
| 1 | slug | meaning |
| 2 | ASAP | As Soon As Possible |
| 3 | BF | Boyfriend |
| 4 | bravo | brave |
| 5 | FF | Friends Forever |
| 6 | GF | Girlfriend |
| 7 | HNY | Happy New Year |
| 8 | HUH | what |
| 9 | ICYMI | In Case You Missed It |
| 10 | IYKWIM | If You Know What I Mean |
| 11 | LG | Local Government |
| 12 | LULZ | Laughs |
| 13 | NP | No Problem |
| 14 | OBE | Overcome By Events |
| 15 | OMG | Oh My God |
| 16 | PSA | Public Service Announcement |
| 17 | QOTD | Quote Of The Day |
| 18 | ROFL | Rolling On Floor Laughing |
| 19 | ROTFL | Rolling On The Floor Laughing |
| 20 | TC | Take Care |
| 21 | U | You |
| 22 | VIP | Very Important Person |
| 23 | W8 | Wait |

Table 4.3: Comparison of Sentiments detection result of Synesketch library and Stanford CoreNLP

| Tweet | Synesketch | CoreNLP |
|---|---|---|
| RT @Aimalkhankakar: Imran Khan, I'll support you till death. I know you will always make us all proud like today. Respect your decision c | Negative | Very Positive |
| @AslamChandio You want thrill and skills. Hmm smart (impatient) guy. | Negative | Very Positive |
| RT @faisalahmadj: @AnsariAdil Na-maloom ko aag lagao, Yeh link khoob share karao! Imran wants govt to prosecute Zahra Shahids killers htt | Negative | Very Positive |
| RT @G0 NAWAZ G0: Congratulations Pakistan Pakistani Players looked Soooo Excited, Happy, Angry Sexy, Crazy celebrating After winning | Negative | Very Positive |
| Today I spent half the day with great #Pakistanis who work smartly for #Kashmir's freedom, end to #Indian occupation, and lasting peace. | Negative | Very Positive |
| RT @MahaKhalidPTI: With @DrAwab @SadiaAgha @DrSeemaSZia at #Karachi . A day dedicated for our people with special abilities! | Negative | Very Positive |
| #Malala,17 accepts #NobelPeacePrize, gives a powerful speech, making the rest of us proud to be #Pakistanis. What a symbol of pride & hope. | Negative | Very Positive |
| @shahzadrez khan, pathan, Tufaan is always unpredictable! | Negative | Very Positive |
| Shanaz Sheikh and management deserves a lot of applause and credit for transforming our hockey men to a potent team! #PHF | Negative | Very Positive |
| @ImranSaeedKhan1 Arrray .. I know one Jiyala, behind the red one, very well.. the other, on the right is I guess, Jamil Soomo | Negative | Very Positive |
| RT @BinaShah: So proud to see Malala in Oslo and Kainat and Shazia, also shot with her, in the audience. Pakistani girls are the pride of | Negative | Very Positive |
| RT @Faraankhan: The biggest positive from the marches is that people now know their rights,their right to live, their right to vote! #GoNaw | Negative | Very Positive |
| Really,I salute azadi March and Inqilab March, they are making history ! #AzadiMarchPTI #inqilabmarch | Negative | Very Positive |
| Well done Roze News, Metro News, Capital TV and Jaag on refusing to air Imran Khan's daily rant. | Positive | Very Negative |
| Also Bilawal, like Imran, sounds bitter, divisive, derisive—unappealing strategy in a fragmented environment. | Positive | Very Negative |
| police, Lower Judiciary, Revenue collection, corruption, merit policies & good governance etc can be improved without foreign trips of PM. | Positive | Very Negative |
| Really disappointing to see PTI supporters are bashing Javed Hashmi and admiring politician like Sheikh Rasheed, Im sure IK wont like it. | Positive | Very Negative |
| RT @BlackParado: So, @HniaziISF justifying attack on @SanaMerza says it is "right", He shouldn't cry if PTI women are trashed by PMLN. | Positive | Very Negative |
| Federal Propaganda Minister repeats 'empty chair' song, says on Beeper, 9000 chairs arranged, half are empty in Islamabad PTI Jalsa :)))) | Positive | Very Negative |
| Living in luxurious house, driving in Cruiser, proper food, coke, not ready to sit in dherna, sleep on container, this is IK ka naya Pak | Positive | Very Negative |
| RT @fasiranjha: PTV, Parliament House, President House, PM House, Geo office got attacked but thank God, 'Gamla' remained safe! | Positive | Very Negative |
| GHQ/politician problem common: they pretend to know all and are too arro-gant to learn, too lazy to inquire! | Positive | Very Negative |
| RT @anum5: Gullu Butt with Khalid Butt and TUQ (parody) too good . Watch #MediaAzadHa for a good laughter #ExpressNews #Pakistan | Positive | Very Negative |

Table 4.4: Event wise tweets with journalist's participation

| **Event** | **Tweets** | **Participating** | **Journalists in percentage** |
|---|---|---|---|
| Zarb-e-Azab | 11,007 | 605 | 65.55% |
| Azadi and Inqlab March | 144,845 | 796 | 86.24% |
| Hockey Champions Trophy for Men 2014 | 13,222 | 597 | 64.68% |

Table 4.5: Event wise sentiment classification with CoreNLP and Synesketch

| **Event** | **CoreNLP** | | | | | **Synesketch** | | |
|---|---|---|---|---|---|---|---|---|
| | Very Negative | Negative | Neutral | Positive | Very Positive | Negative | Neutral | Positive |
| Zarb-e-Azab | 42 | 8,309 | 2,061 | 592 | 3 | 7,416 | 2,202 | 1,389 |
| Azadi and Inqlab March | 520 | 100,870 | 31,295 | 12,021 | 139 | 85,606 | 32,715 | 26,524 |
| Hockey Champions Trophy for Men 2014 | 25 | 8,588 | 2,929 | 1,651 | 29 | 7,972 | 2,450 | 2,800 |

## 4.7 Results

Table 4.4 shows the event wise tweets and journalists' participation. The data of event related tweets and journalist participation shows that our framework is able to find users of a targeted community, retrieve their tweets, and filter tweets based on events.

To test our first hypothesis that "journalists tweet on events occurring in their country". Table 4.4 shows that in event "Zarb-e-Azab" 65.55% of journalists participated and this is the minimum participation level of all the three events. Because we have filtered events related tweets on the basis of keywords and there is a chance that the event related tweets are not filtered, that do not contain given keywords.

Further for the qualitative analysis, we randomly chose 30 "Zarb-e-Azab" related tweets. These tweets are shown in table 4.6 along with their user ids. As we see in table 4.6 at serial numbers 10 and 24, that the twitter ids looks like any news agency instead of journalists. Because many users keep the journalists ids and news agencies ids in the same twitter list, this affects the process of snowball sampling, and causes some irrelevant users in the targeted community. It has also been observed that sometimes journalists express their opinion in different context, but these tweets are filtered as related to the event because they contain our selected keywords, e.g., tweets in table 4.6 at serial numbers 4, 10, and 17 are not related to the event "Zarb-e-Azab" but these are in dataset due to related keywords.

To measure the effectiveness of the proposed framework, we compute precision and recall. Although precision and recall should be measured on the complete dataset, and as we do not have a complete labeled dataset, therefore we compute precision and recall on a random sample. This section presents the precision and recall levels of various events relevant tweets on a random subset of the complete dataset. Precision and recall of sentiment analyzers are discussed in section 4.8. Table 4.7 shows the precision for all the three events. The precision and recall are computed as follows:

$$Precision = \frac{t_p}{t_p + f_p} \tag{4.1}$$

$$Recall = \frac{t_p}{t_p + f_n} \tag{4.2}$$

where $t_p$ the number of relevant tweets fetched in the random sample, $f_p$ is the number of irrelevant tweets fetched in the random sample, and $f_n$ is the number of relevant tweets not fetched in the random sample.

The top contributor in the "Zarb-e-Azab" event was Mr. Farooq Mahsud[22] (according to twitter profile his location is FATA and description is "#Journalist, #PeacePromoter From #SouthWaziristan Agency, Do or Die is n old concept ,,,,,,Do it Before you Dieper"). He contributed with 561 tweets. Second contributorwas Mr.

[22] https://twitter.com/MahsudFarooq

Safdar Dawar[23] (location is Peshawar Pakistan and description is “Former President Tribal Union of Journalists, FATA Pakistan..Belong to North Waziristan Agency” ) with 241 tweets.

As per table 4.4, 86.42 percent, journalists participated in the “Azadi and Inqlab March” event. Participation in this event was more than the other two events. The reasons could be extensive media coverage and easy access to these marches.

We randomly chose 30 tweets related to the event as shown in table 4.8 along with their owners ids. In table 4.8 at serial no 26 and 27 the twitter ids belong to news agencies and blogs and at serial no 4 and 18 belong to politicians, because some users kept news agencies and politicians in the same twitter list. Mostly politicians are found on these lists who give their party opinion on different issues in the talk shows on news channels. The tweets at serial no 7, 10, and 13 are not relevant to the event and authors expressed their opinions in different context. Based on this sample, the precision is 0.9. The two top journalists who participated in this event were Mr Qaiser Chishty[24] and Mubasher Lucman[25] with 1343 and 980 tweets respectively.

Very low number of users participation in the event “Hockey Champions Trophy for Men 2014” as shown in table 4.4 is due to the limited duration of one week for this event. 64.68 percent journalists expressed their opinions in the context of this event, which is the least participation as compared to the other two events.

For the qualitative analysis, we randomly chose 30 tweets from the filtered dataset as done previously. The table 4.9 shows these 30 random tweets. When we check the profiles of owners of these 30 tweets, we found that owner ids at serial no: 2, 17, 19, and 21 in table 4.9, do not belong to the journalist community. Owner id at serial no 2, “ArifAlvi” belongs to a political personality and owner ids at serial 17, 19, and 21 belong to news agencies. Tweets at serial no 4, 19, 23, 26, and 27 are not relevant with the event “Hockey Champions Trophy for Men 2014”, but these are filtered due to selected keywords. The precision of “Hockey Champions Trophy for Men 2014” event is 0.83.

[23]https://twitter.com/DawarSafdar

[24]https://twitter.com/QaiserChishty

[25]https://twitter.com/Mubashirluc
man

Table 4.6: Random Sample tweets for the event "Zarb-e-Azab"

| S.no | Twitter id | Tweet |
|---|---|---|
| 1 | Aak0 | #ZarbEAzb The other side of peace: scared residents flee the war zone |
| 2 | Adnanrandhawa | Wondering when CIA senate committee members disappear and reports "leaked" to media they have gone to N. Waziristan wilfully. |
| 3 | adilshahzeb | See how good he sounds here MT "@MurtazaGeoNews: TuQ says he'll dispatch 14 truckloads of food/medicine today4 IDPs in Bannu,more2 follow" |
| 4 | ajmaljami | —RT @ReesEdward: The British in N. Waziristan. Sometime in early 20th century. |
| 5 | alisalmanalvi | Dear COAS, the killers of these innocent school children are not restricted to North Waziristan only. They are everywhere in #Pakistan. |
| 6 | AmirMateen2 | need to something fast What started as a mass exodus of locals is now humanitarian crisis" http://t.co/rokEZDExKz Waziristan @DalrympleWill |
| 7 | madihariaz | Samaa ran a report mocking the cricket team about the amount of their donation for IDPs. Then asked a moulvi's opinion who deplored it too |
| 8 | MahsudFarooq | Effect of Terrorism on music in South waziristan . @AnserAbbas @alex gilchrist @FATANews @IftikharFirdous @pirroshan |
| 9 | MahsudFarooq | Clash b/w security forces and millitants, 5 Millitants killed in sarvakai area #south #waziristan agency, security sources |
| 10 | ApnaWaziristan | New Delhi: Kashmir Bharat Ka Atoot Hissa Tha, Hissa Hein Aur Rahay Ga, Pakistan Ko Sirf Apni Fikar Karni Chahiye:#BJP |
| 11 | ApnaWaziristan | *Peshawar: Kpk Hakumat Ny IDPs Ky Liye Shikayat Cell Qaim Kr Diya, IDPs Peer Sy Hafta Subha 9 Bajay Tak Shikayat Darj Kara Saktay Hyn: |
| 12 | AtikaRehman | RT @TahaSSiddiqui: Since we've SO MUCH of aid coming in for #Waziristan IDPs, why not waste some thru poor logistics and arrangements! http |
| 13 | BenazirMirSamad | RT @PTIofficial: Khyber Pakhtunkhwa Govt making adequate arrangements for IDPs. Instructions passed to all relevant departments. |
| 14 | SaeedShah | RT @asadmunir38: Suicide attack kills four soldiers in North Waziristan http://t.co/eGRQNMpXBT |
| 15 | DawarSafdar | Afghan gov will teach IDPs children,and here KP Gov ordered to vacate the schools |
| 16 | DawarSafdar | In thousands IDPs in government schools going to IDPs again. |
| 17 | FarooqHKhan | RT @Shahidmasooddr: We pay taxes to Govt for IDPs/flood victims etc.We dont pay taxes to opposition! And its better to beg than to Rob jana |
| 18 | FauziaKasuri | RT @imran sidra: Ma'am @FauziaKasuri & team while doing clothes shopping for IDPs sisters in Bannu KP. #HelpIDPs #Donate #IKF #PTI http://t |
| 19 | FauziaKasuri | @ArshadSidiqi Thank u for thinking of the IDPs..Allah bless you all. |
| 20 | taahir khan | @Jan Achakzai JUI-F should start protest to force the gov't, military to send back 1 million North Waziristan people instead political Jirga |
| 21 | MinaSohail | RT @asadmunir38: #PakArmy soldier #ZarbeAzb |
| 22 | P Musharraf | I vehemently condemn the suicide attack on our troops in North Waziristan. The ultimate sacrifice offered by our... http://t.co/nM4MXtXcwt |
| 23 | PakMilitaryNews | RT @AsimBajwaISPR: #ZarbeAzb:A pic taken in #IDP Camp Bannu today. Let us join hands to bring back their smiles#helpidps |
| 24 | PakMilitaryNews | Pakistan plans military operation in North Waziristan, targeting extremist groups |
| 25 | KlasraRauf | RT @Dr Afaq: @KlasraRauf doesn't talk of utopia in his Urdu column.He talks about real life solution to #IDPs problem. |
| 26 | penpricker | For me those nameless innocent kids of Waziristan, who die in drone strikes r no less than Malala. All of 'em are victim of a war not ours |
| 27 | muniraqazi | @dunyanetwork @BBhuttoZardari Good one! The PPP's support for #ZarbEAzb is vital for a stable & secure #Pakistan. #PPP |
| 28 | NadiaaQasim | 4ñ£lrt¸ 4atr ¸va s,s ,Z ¿ı,ítla .tlr,ı,9 ,st¸v ç9 ,t 4ll£l3¸ ,S ¸k, ç,t¥ ,mu - 9lZ" ,Z ¿ı,ítl9¸t¸lk, ø9S ,ña ,Z ¿a9 ¿ñ£9nj,í #IDPs #Bannu |
| 29 | nadia a mirza | No Federal everything trickles down to army to serve n manage #IDP's, neither Provincial Govt presence, only photo sessions. |
| 30 | QuatrinaHosain | RT @Khawar69: Media shud rather demonise the ideology of TTP and Jandullah that preach killing of humans for political goal. #zarbeazb will |

Table 4.7: Tweets relevancy with events

| Event | Precision |
|---|---|
| Zarb-e-Azab | 0.9 |
| Azadi and Inqlab March | 0.9 |
| Hockey Champions Trophy for Men 2014 | 0.83 |

The two top contributors were Faizan Lakhanih[26] (location is Karachi, Pakistan and description is “Journalist based in Karachi. Sports Reporter - Geo News. Tweets are personal opinion”) and Mehr Tarar[27] (Location is Lahore, Pakistan and description is “His mom. Former Op-ed Editor (Daily Times, Pakistan). Columnist. Veritas vos liberabit. Words matter. ...as certain as the sun rising in the east... ”) with 242 and 221 tweets respectively.

**Measuring Recall:-** It is straight forward to compute precision on a random sample, but recall cannot be computed unless complete dataset is labeled. As we do not have a labeled dataset, to measure recall, we randomly choose 50 tweets for each of the three events within the event duration but without using any keywords (i.e., the tweets not fetched by the framework). If the recall of our framework is 100%, the randomly chosen tweets should not contain any relevant tweet (because those should be fetched by the framework). On the contrary, the sampled tweets contain a few relevant tweets. Out of each set of 50 random tweets, 1, 7, and 2 tweets were relevant (false negatives) to “Zarb-e-Azab”, “Azadi and Inqlab March”, and “Hockey Champions Trophy for Men 2014” events respectively. The recall is measured as 0.98, 0.86, and 0.96 for events “Zarb-e-Azab” “Azadi and Inqlab March”, and “Hockey Champions Trophy for Men 2014” respectively. It is possible that the sample of 50 tweets for the each event was not large enough for such a large dataset. On the other

[26] https://twitter.com/faizanlakhani

[27] https://twitter.com/MehrTarar

Table 4.8: Random Sample tweets for the event "Azadi and Inqlab March"

| S.no | Twitter id | Tweet |
|---|---|---|
| 1 | abdullasyed | RT @iamkhiofficial: "I'm overwhelmed by the number of people & their passion for participation."- Chris at #Iamkhi youth festival. http://t |
| 2 | ajmaljami | @mkw72 Chicken or beef? Hallaaal?? @Fooka Online @ImranKhanPTI |
| 3 | ammarmasood3 | RT @MianVirgo: @asifiatislamabad @ammarmasood3 @ImranKhanPTI [illegible] #GoNawazGo |
| 4 | ArifAlvi | @tariqakbari @ImranKhanPTI @arsalan muneer Turk system is not backed by the army at all thanks to Erdogan and his AKparty |
| 5 | MaheenUsmani | While Khan Sb eats/sleeps in Bani Gala mansion,wet & hungry #PTI men await Inquilaab @MurtazaGeoNews http://t.co/I41BhX1Qai #AzaadiMarchPTI |
| 6 | MariaMahesar | RT @InsafRadio: Imran Khan addressing to huge crowd at #AzadiSquare on 45th day of Azadi March. Listen Live http://t.co/phftiIjttu http:// |
| 7 | bamalik123 | @arsched #Politics like #patriotism has become refuge of the scoundrel |
| 8 | DrDanish5 | Is there any plan in the Camp of Imran Khan or Dr Tahirul Qadri ??? Kub tak esi tarah bethein ge? #AzadiMarchPTI #InqilabMarch |
| 9 | BeenishSaleem | RT @AnjumContact: @BeenishSaleem love it wn u call @TahirulQadri "madrassay wali sarkar".Hats off to u.Doing excellent job during scripted |
| 10 | SaifanKhan | RT @firetree101: @SaifanKhan Gaza |
| 11 | ShafiNaqiJamie | Final dead line and Dharnay. Govt and strategy Haroon Rashid gives you over view. Sairbeen is on air |
| 12 | shahbazzahid | RT @AapaZubeda: Mujra option bhi rakh detay beta sawal mai @Snobish 02: @Aa- paZubeda Question: Aapa kia Ajj ka Jalsa Waqai me Jalsa tha ya |
| 13 | shahzadrez | I did it as well. Never knew trolls had audacity of defaming patriotic armed forces RT @mughalbha Blocked @Huzefa1983 @AsimBajwaISPR |
| 14 | TalatHussain12 | Mind-boggling backtracking. In interviews, pressers and speeches Khan described CJ Iftikhar an alleged lord of rigging mafia. |
| 15 | hinaparvezbutt | RT @dunyanetwork: Lahore traders reject PTI's call for strike http://t.co/Nx1kXHLdAk http://t.co/bsnQAhJzfm |
| 16 | wajih sani | RT @Majid Agha: Imran Khan raises fingers on everyone, & claims "false" high moral grounds. 11 #PTI MNA doesn't pay Tax: FBR Report |
| 17 | WasseemBadami | RT @ZaidZamanHamid: If army cannot help the people in getting justice, then it must step aside also & let the revolution takes its course ! |
| 18 | NazBalochPTI | RT @arfa rafiq: @NazBalochPTI Happy birthday Milady. a dazzling guidance for women in #Pakistan and yes #azadi Icon .May u've many more dea |
| 19 | nazsahito | RT @SoomroJameel: @ImranKhanPTI plz check your #KPK govt. are they imple- menting #specialPerson qouta in govt. Jobs. Which started in #SMBB |
| 20 | kalsoom82 | RT @mosharrafzaidi: Corruption is terrible. But the last few days teach us that of the seven deadly sins, none is as lethal as vanity. |
| 21 | kashifsabir | The scenes of red zone after #TUQ sit in #Islamabad#Pakistan http://t.co/Tpp0zWRVgq |
| 22 | kashifsabir | RT @MurtazaGeoNews: Two months after #PTI #PAT raided Islamabad to attempt a bloody coup, the police takes control of the Constitutional Av |
| 23 | mnyaseen | RT @FatimahLove92: #InqilabMarchWithDrQadri - Updates by Mr @aliimumtaz —#TuQ http://t.co/XcYYAiPagU |
| 24 | mohsinrz | RT @YusraSAskari: Almost 100 dead in torrential rains across the country and never ending container top rants are still national focus - #O |
| 25 | MuhamadAfzalECP | RT @ShkhRasheed: Good to see janoon at star gate dharna. |
| 26 | PKMediaNews | RT @SMQureshiPTI: The negotiation team on both sides agreed on all other points except for the 1 whereby v demanded resignation of PM till |
| 27 | pkpolitics | RT @sanabucha: Resignations? PTI in an effort to prove 'we mean business' by going 'out of business'! Only 'business as usual' in KPK. Vah! |
| 28 | muneebfaruq | @NavidAhmadKhan Think through this. 50m ballots.. Where is the proof? Evidence? EC rejects.Not bothered about Govt. it will go if its fake. |
| 29 | QasimNauman | RT @MaryamNSharif: Ice bucket challenge! RT @beenasarwar: Good grief Pak RT @javerias: Another view of #parliament aftr AzadiMarchPTI htt |
| 30 | qayyum nizami | @AQKKundi inshallah u will lead revolution one day |

hand it assures that the mechanism of filtering tweets on the basis of keywords is effective.

## 4.8 Journalist Opinion and Sentiment Analysis

To avoid the biasness of the sentiment analyzer, we used two sentiment detection tools, Stanford CoreNLP and synesketch. Stanford CoreNLP gives more accurate results than the synesketch tool in situations where phrases contain positive as well as negative words, e.g. tweet from table 4.3 “Living in luxurious house, driving in Cruiser, proper food, coke, not ready to sit in dherna, sleep on container, this is IK ka naya Pak” was classified as positive by synesketch and very negative by CoreNLP. When in the phrase there are complete sad words with happy emoticons, the sysnesketch mark it as a positive sentence, because it first searches for emoticons in the phrase, if it finds emoticons then it marks the sentiments of phrase on the basis of emoticons and do not check it further. However the CoreNLP analyzer checks the phrase in every aspect (i.e. words polarity, emoticons and grammatical rules) and then assign the sentiments.

As another tweet from table 4.3 “Federal Propaganda Minister repeats ’empty chair’ song, says on Beeper, 9000 chairs arranged, half are empty in Islamabad PTI Jalsa :)))) ”classified as positive by synesketch due to present of smiley emoticons and very negative by CoreNLP due to semantic and syntactic structure of tweet. It is important to compute the precision of both the sentiment analyzers (Synesketch and Core NLP). This will help us in quantitatively analyzing the analyzers as well as help us for choosing appropriate sentiment analyzer for further analyses.

To calculate the precision of sentiment analyzers, we choose thirty tweets randomly from each event, as we did for the precision of event relevant tweets in section 4.7. The tables 4.11, 4.12 and 4.13 show randomly selected tweets of the three events “operation

Table 4.9: Random sample for the event "Hockey Champions Trophy for Men 2014"

| S.no | Twitter id | Tweet |
|---|---|---|
| 1 | MurtazaGeoNews | RT @usmanmanzoor: It took 20 years for Pakistan to reach Champions Trophy Final; All players deserve good jobs in PSEs for financial support |
| 2 | ArifAlvi | RT @faizanlakhani: Pakistan Hockey players blow flying kisses to spectators at sta- dium, I think they've sent the clear message "forget what |
| 3 | SaadiaAfzaal | RT @ZaidZamanHamid: Pakistan has won the Champions trophy already by crushing India ! Now it does not matter even if Germany takes the cup |
| 4 | faizanlakhani | Quaid e Azam Trophy: Match between WAPDA & SNGPL at SouthEnd Club com- pleted within two days, four innings in 147.4 overs. #Cricket #Pakistan |
| 5 | AqilSajjad | RT @mak asif: Reminder to all: These hockey players are jobless, haven't had central contract money for 18 months, were sponsored to CT by |
| 6 | anumuae | Geo tu Aasay Waqyee or who bi India Mein #PakvsInd #ChampionsTrophy Shabash Meray Jawan Proud of You http://t.co/kvm4MTbdPS |
| 7 | AsmatullahNiazi | congratulations #Indian #Media for success in getting ban from #FIH is it not a biased decisions #fihockey |
| 8 | khalidkhan787 | RT @azfar25: Two Pakistani players suspended for one match by FIH. FIH you pissed & India you are bad #Loosers |
| 9 | hamdani raza | Another defeat ... Australia win bronze medal #hockeychampionstrophy |
| 10 | Aoon Syed | A Simple Question To All Indians....Hamary players Ne shirts Utar Lien to Masla Hogea ...Khud Gandhiiii Ne Sari Zindgi Shirt Pehni Kla |
| 11 | farzana versey | Mature? Those shrivelling over 'insult' to nationhood by some finger-showing on the hockey field get back by showing Kargil fist. #INDvsPAK |
| 12 | madversity | Viren is good at bullying off. Suhel Seth is good at bullying. Viren is good at hockey. Suhel is good at hocus pocus. Small differences |
| 13 | KiranKhanLive | RT @MianZohaib : We are proud of you boys in green. #BleedGreen #PakvsGer #ChampionsTrophy #RoIndiaRo http://t.co/EJuFQVEr3I |
| 14 | mirza9 | RT @barcastuff: Barcelona are the first team in history to win their Champions League group in 8 consecutive seasons #fcblive |
| 15 | HaroonRsh | RT @TalatHussain12: Banning Pakistan hockey players is unjust. Why isnt this our media's biggest story? |
| 16 | rehman azhar | In @FollowUpRehman today, we are joined by Rehan Butt(PAK) and Shekar Louthra (IND) on hockey controversy at 7:03 pm on #DunyaTV. |
| 17 | mediagag | Waise it would be amazing if Pakistan did win one of the three major titles just as the Waseem/Sohail generation finally left |
| 18 | AnserAbbas | Two players banned from tonight's Champions Trophy final despite Pakistan apology to India http://t.co/mElmoFx1bw via @etribune |
| 19 | PakPassion | #PakvNZ Pakistan win toss and have decided to bat in the 2nd ODI in Sharjah - teams to follow.... |
| 20 | khalidkhan787 | RT @KhanDanish : 21 middle-finger salute to @FIH Hockey for their utter bias. #ShameOnFIH |
| 21 | SportsReporter | Champions Trophy Hockey : Pakistan lost to Belgium by 2-1 . |
| 22 | NasimZehra | Inability of khakiFactor 2 make substantive impact on fortunes of pol players post-Aug |
| | | fairly clear-only pol own sins/virtues wil help/hurt |
| 23 | ninoqazi | finally winter has arrived in #Delhi. overcast with a little wind & drizzle |
| 24 | aZmAtNaWaZmAli K | RT @MalihaMansoori: Virat Kohli shows Finger to #Australian crowd and gets<br>Fined. #Pakistan #Hockey Players Celebrate their Win and they  ge |
| 25 | AnsHafeez | Nara e Takbeer Allah Hu Akbar shouted At Indian Ground... Pakistan Zindabad<br>#Hockey #PakvInd |
| 26 | khalidkhan787 | pic of wall collapse in Huaibei, China, in which 5 female players killed http://t.co/LRgH8MKnn0 |
| 27 | myraemacdonald | (Who wins the escalation if #India starts targeted killings in #Pakistan?) India came<br>close to killing Dawood in Khi http://t.co/8dEF0xUOx3 |
| 28 | mosharrafzaidi | RT @najamsethi:  FIH has just cleared all Pak Hockey players of any wrongdoing<br>after aassurance that players meant no offence by joyous cele |
| 29 | SportsReporter | Champions Trophy Hockey : England out class Pakistan by 8-2 . |
| 30 | MehrTarar | RT @JasiJasia: To all those who are criticizing Pakistani hockey players for their<br>obscene gestures after winning against India http://t.co |

Zarbe-e-Azab" , "Azadi and Inqlab March" and "Hockey Champions Trophy for Men 2014" respectively . In these tables each tweet is also manually labeled. We use equation 4.1 for computing the precision, $t_P$ is the number of tweets with correctly identified sentiments and $f_P$ is the number of incorrectly identified sentiments. The tweets with incorrect sentiments are highlighted in these tables. Table 4.10 shows the precision of Synesketch and Stanford CoreNLP. As expected, the precision of Stanford CoreNLP is much higher than Synesketch for all the events.

Table 4.10: Precision of Sentiment Analyzers

| Event | CoreNLP | Synesketch |
|---|---|---|
| Zarb-e-Azab | 0.8 | 0.7 |
| Azadi and Inqlab March | 0.83 | 0.73 |
| Hockey Champions Trophy for Men 2014 | 0.73 | 0.46 |

Therefore to address our hypotheses we will refer the results of Stanford CoreNLP sentiment analyzer

Our second hypothesis (section 1.3), states that a community gives its opinion in tweets on events related to its country. The pie charts in figure 4.1 shows the sentiment wise journalists results. Figure 4.1a shows the sentiment cluster of the event "Zarb-e-Azab" in which 32% of journalists, sentiments belongs to a single category (e.g. positive or negative), 34% of journalist express their opinion in two categories (e.g. positive and negative), 30% of journalists sentiments belongs to three categories (e.g. positive, negative and neutral), only 4% of Journalists express opinion in four categories (e.g. positive, very positive, negative and neutral) , and in this event no journalist is found, who expresses his opinion in all the five categories (i.e. very negative, negative, neutral, positive and very positive).

Table 4.5 shows that mostly the journalists express their opinion negatively because they are talking about the terrorist activities, IDPs problems and also express opinions against the "Zarb-e-Azab" operation, E.g. in tweets at serial no 1 (" #Pakistan rulers claims of #ZarbeAzb n #Waziristan exposed wth #PeshawarAttack.Fighting paid US War of Terror is bringing mayhem inside Pak. " ) and 7

Table 4.11: Random sample tweets from the event Zarbe-e-Azab with analyzers and manual sentiments

| S.No | Twitter id | Tweet | Synesketch | CoreNLP | Manually Tagged |
|---|---|---|---|---|---|
| 1 | baqirsajjad | Good riddance! @AroojZahra Just in: 8 killed several injured as two groups of Taliban open fire at each other in Shaktui, South Waziristan | Negative | Negative | Negative |
| 2 | khushnood2020 | RT @AQpk: This #Pakistan hero died clearing #TTP landmines so that our #IDPs families can return to a clean area. #ZarbEAzb http://t.co/F3 | Negative | Positive | Positive |
| 3 | MudassarGEO | RT @NazranaYusufzai: What would be the meeting point of #drones and #Fazluallah - would he go to Waziristan or drone would come to swat. | Negative | Negative | Negative |
| 4 | AdeelHashmi3 | RT @ImranKhanPTI: I addressed the NWA Jirga which appreciated PTI's position on & support 4 IDPs.The same Jirga had refused to meet PM. htt | Negative | Negative | Negative |
| 5 | taahir khan | I have seen only Islamic charity groups like Jamaat-e-Islami-linked Al-Khidmat and Flah-e-Insanyat engaged in IDPs relief in Bannu | Negative | Negative | Positive |
| 6 | humayusuf | RT @AhmadShuja: South Waziristan's last girl's secondary school blown up: http://t.co/14cffeja | Negative | Neutral | Negative |
| 7 | Khalid Aziz | RT @MunizaeJahangir: A beautiful well researched article on the war torn Waziristan by @OwaisTohid its poetic http://t.co/VqvullhXE5 | Negative | Positive | Positive |
| 8 | penpricker | RT @Uqab_: Had #ZarbEAzb been US backed then y would US say that thy were not informed http://t.co/zAh0Bf41N2 @kashmirrebel @naqkash | Negative | Negative | Negative |
| 9 | Farwa_Chaudhry | RT @IftikharFirdous: 3 security personnel killed in in North Waziristan 10 injured, 4 killed in South Waziristan, 12 injured, 16 killed in | Negative | Negative | Negative |
| 10 | DawarSafdar | RT @TahaSSiddiqui: Op Zarb-e-Azb: Residents of N Waziris-tan given one more day to evacuate http://t.co/yjZP80JadZ http://t.co/0FMdNxxzug | Negative | Negative | Negative |
| 11 | alishahjourno | RT @smstahir: Bring back their smiles #IDPs #ZarbEAzb @mur-tazasolangi @beenasarwar @asmashirazi @AsimBajwaISPR @PakMili-taryNews http://t.co | Negative | Positive | Positive |
| 12 | Bitani1 | Infrastructure worth Rs1b turned into rubble in N. Waziristan http://t.co/EZCxwEHhfD | Neutral | Positive | Negative |
| 13 | moni_butterfly | It's come in Wicked Leaks that army was asking America for drawn plane strikes on Waziristan. Haw, hippo crates jaisay | Negative | Negative | Negative |
| 14 | jasidiqi | #PTIpeacemarch allowed to proceed to South Waziristan, claims @Shafqat_Mahmood . | Neutral | Neutral | Neutral |
| 15 | zaighamkhan | RT @OwaisTohid: Do remember 1/2 million displaced tribesmen of N.Waziristan in Ramazan. Donate,embrace them&reject extremists #IDPs http:/ | Negative | Negative | Negative |
| 16 | Khadimhussain4 | Waziristan IDPs finding it hard to feed their cattle http://t.co/VRDPNfAm7e | Negative | Negative | Negative |
| 17 | NadiaaQasim | RT @CapitalTV News: Operation #ZarbeAzb: IDPs facing countless problems, for details watch #CapitalPoint now on #CapitalTV. | Neutral | Negative | Negative |
| 18 | asadmunir38 | Baitullah Mahsud was an imam masjid of a small mosque in FR Bannu,came to South Waziristan in 2002,fled to North in 2004 with only 20 men. | Negative | Negative | Negative |
| 19 | shaistaAziz | ISPR now saying 30 militants killed in shelling by helicopter gunships in North Waziristan: http://t.co/Ik1Ww53vFk #Pakistan via @etribune | Negative | Negative | Negative |
| 20 | SAfridiOfficial | Alhamdullilah 1st step for the IDPs and we will support them more & more inshAllah http://t.co/osE0ePdoLh | Neutral | Negative | Positive |
| 21 | Nabeehae | Number of IDPs reaches 5 lakh 72 thousand http://t.co/4BTghkADmJ #Pakistan #Karachi" | Negative | Negative | Negative |
| 22 | taahir_khan | Spent hours with Waziristan displaced persons at Bannu. On their first day of Eid IDPs remembered previous Eids at homes. They want 2 return | Negative | Negative | Negative |
| 23 | Khalid Munir | #ZarbEAzb This is the beginning of the end of terrorism in Pakistan, DG ISPR http://t.co/pjQpqBXRQq | Negative | Negative | Positive |
| 24 | aliarqam | Now #IDPs are beaten by #Police in #Bannu people from #PPP #PMLN who closed doors of their provinces on IDPs are shedding crocodile tears | Negative | Neutral | Negative |
| 25 | AtifSal | #Pakistan fighting #ZarbEazb, implementing National Ac-tion Plan 2 secure #US, Indian interests. Our rulers love $$ http://t.co/VOAPky6BWu | Positive | Negative | Negative |
| 26 | fareedraees | @ImranKhanPTI @PMNawazSharif 2 #Drone attack in North Waziris-tan. 5 dead and three injured. An open question for #PMLN & #PTI | Negative | Negative | Negative |
| 27 | nighatdad | God save them from PPA re cyber crimes. RT @The_Nation: 16000 IDPs provided mobile SIMS: http://t.co/81ITBVunm5 http://t.co/deeT4LTcoY | Negative | Negative | Negative |
| 28 | ShaiziCheema | RT @alisalmanalvi: #PTI endorses Zarb-e-Azb operation. Just a week ago IK said that such an operation would be 'suicidal'. http://t.co/BKAc | Negative | Negative | Negative |
| 29 | AliDayan | Time for detailed information on 376 terrorists killed in the military op. Also for free media access to 19 detained. #Pakistan #Zarb-e-Azb | Negative | Negative | Negative |
| 30 | paras_jn | RT @BBhuttoZardari: To all the martyred soldiers Zarb-e-Azb & other great battles we salute you for your sacrifices & the nation stands beh | Positive | Positive | Positive |

Table 4.12: Random sample tweets from the event Azadi and Inqlab March with analyzers and manual sentiments

| S.No | Twitter id | Tweet | Synesketch | CoreNLP | Manually Tagged |
|---|---|---|---|---|---|
| 1 | vogul1960 | Shahbaz Sharif has to find an acceptable way out on the killing of 20 plus PAT workers on June 17. No escape from this, it seems! | Negative | Negative | Negative |
| 2 | Goshno | So what have the Inqilaabis achieved? People the youth are more interested in politics, parliamentary sessions (joint) are active | Negative | Neutral | Neutral |
| 3 | Senator_Baloch | RT @omar_quraishi: Surpeme Irony — PTI wants free and fair elections but is not directing its march at Aabpara – and in fact is being g | Negative | Negative | Negative |
| 4 | nighatdad | @emrys s Digital Political party - Glad that PTI seriously consider these spaces. Have they set a good precedent? your thoughts? | Positive | Negative | Negative |
| 5 | ayeshaijazkhan | @FKhan145 btw I find team leader determined and most dishonest. | Negative | Negative | Negative |
| 6 | TenzilaMazhar | @PTIofficial again and imp speech.Well still hope for sum concrete announcements coming up from khan,integrated vision.A well thought speech | Negative | Positive | Positive |
| 7 | KabirWasti | Had there been a genuine accountability in Pakistan, people like Zardari and NS would been out of politics. #PTI @arsched @SSEHBAI1 @ZeeNews | Negative | Negative | Negative |
| 8 | aamirghauri | RT @SyedIHusain: I am seriously feeling sad for Mr Khan :( | Negative | Negative | Negative |
| 9 | AtikaRehman | RT @umairjav: Qadri sb, hope that bread you're eating is brown. Cuz otherwise it'll take an inquilab to make your bowels move. | Negative | Negative | Negative |
| 10 | nausheenyusuf | Plan C is already turning to b Plan D as Deaths are reported in Ambulances due to PTI Blockade in LAHORE.. | Negative | Neutral | Negative |
| 11 | guldaar | @SamadK PMLN khawateen vs. PAT/PTI khawateen. The stereotyping from both sides deserves a research project! | Negative | Negative | Negative |
| 12 | wajih_sani | RT @ovaisjafar: So #ARY reporting #PTI supporters were egged by #PMLN supporters in Faisalabad | Negative | Negative | Negative |
| 13 | mubasherlucman | RT @FaisalYaseenPTI: "Why is this man banned for life from Pakistani TV? http://t.co/HCVvYcnFbS http://t.co/kCC22XPfMv" fr telling th truth | Negative | Negative | Negative |
| 14 | Aoon_Syed | @AnooshNj itna b old nhe hy , kuch month pehlay k hy , when he was keen to join Geo, was in contact with MSR @FarhanKVirk @PTI_Ki_Tigress | Negative | Negative | Negative |
| 15 | MurtazaGeoNews | Metro bus being attacked by 'peaceful' protestors. Imran khan will not condemn them, like he never condemned Taleban http://t.co/IGOoO6hARI | Negative | Negative | Negative |
| 16 | mahvishahmad | RT @monadarling: PTV anchor Uzma Chaudhry, almost in tears, says she was hit with a stick, bag and dupatta snatched by protesters when she | Negative | Negative | Negative |
| 17 | kazmiwajahat | That moment when the national anthem is played before the address of Prime Minister to the nation #AzadiMarchPTI #PAT http://t.co/ogm3E2iPWe | Negative | Neutral | Neutral |
| 18 | TalatHussain12 | Imran's charge-sheet relies on personal testimonies.Will these people speak and furnish evidence? | Negative | Neutral | Negative |
| 19 | MaizaHameed | RT @VoPMLN: PTI flag can be used for dumping used pampers... | Negative | Negative | Negative |
| 20 | peshavar | @KKhanMarwat MAPs cannot come due to other responsibilities, must facilitate people from their respective constituencies to #AzadiMarch | Negative | Negative | Positive |
| 21 | SyedAliHaider13 | @ShkhRasheed nd @TahirulQadri said tht in the end of August govt will be no more! And the same is said by Hashmi against #IK | Negative | Negative | Negative |
| 22 | iamTribalKhan | @ImranKhanPTI Review your decision of marching red zone. | Neutral | Neutral | Neutral |
| 23 | AnsarAAbbasi | Sh Rashid now says he would set the country on fire if Imran Khan is arrested? Earlier it was IK's cousin TuQ now SR inciting violence. | Negative | Neutral | Negative |
| 24 | Nadia_Zaffar | #TuQ should return to Canada immediately - Burger king is taking over Tim Hortons. Culture bhrasht ho raha hai! | Negative | Negative | Negative |
| 25 | najamsethi | RT @Jamil20111Jamil: @najamsethi today ik and Dr shahid masood will say sethi did puncture in Aus team we will do dharna | Negative | Negative | Negative |
| 26 | zuberishahab | RT @fawadchaudhry: I am short of words in condemning ill worded,stupid disgraceful statement of FazaUrRehman he has nt humiliated only PTI | Negative | Very Negative | Negative |
| 27 | KabirWasti | Chief of Pakistan Army Gen.Raheel Sharif will go US on an important visit in November 2014.To me this visit would change many things. #PTI | Negative | Negative | Negative |
| 28 | mehreenzahra | Hey @BBhuttoZardari you watching #ImranKhan 's #Larkana rally? | Neutral | Neutral | Negative |
| 29 | MaizaHameed | RT @MaryamNSharif: Shame Imran Shame ! http://t.co/UD8JD9cMCE | Neutral | Negative | Negative |
| 30 | DrDanish5 | Imran Khan to root out the evil of Injustice poverty slavery ignorance corruption Rulers made the masses their slaves. ......... | Negative | Negative | Positive |

Table 4.13: Random sample tweets from the event Hockey Champions Trophy for Men 2014 with analyzers and manual sentiments

| S.No | Twitter id | Tweet | Synesketch | CoreNlp | Manually Tagged |
|---|---|---|---|---|---|
| 1 | alisalmanalvi | Keep calm and cherish this win. #ChampionsTrophy #Hockey #PakvsInd | Neutral | Neutral | Positive |
| 2 | hafsa adil | RT @SalmanAkbar12: Proud of u Pak Hockey team u guys are true soldiers. In these circumstances what u guys have done can't be describe in w | Negative | Negative | Positive |
| 3 | rasheedshakoor | @EshalMirza what you expect from this pakistani team . we have to accept the fact pakistan hockey is now finished . | Negative | Neutral | Negative |
| 4 | IftikharFirdous | RT @khalidkhan787: pics: Celebrations by Pakistan Hockey team in Bhubaneswar, after beating India in India by 4-3 in Champions trophy http: | Positive | Negative | Neutral |
| 5 | Wabbasi007 | Yesssss"@nadiakhaan: Matches either Cricket or Hockey, b/w India & Pakistan, prove one thing clearly- 2 Nation Theory was absolutely right!" | Negative | Negative | Negative |
| 6 | MurtazaGeoNews | Pak TVs should have led the protest against banning of hockey players by #FIH but it remains obsesed with the politics of strikes/lock-downs | Negative | Negative | Negative |
| 7 | usmanmanzoor | RT @NadeemfParacha: In a classic display of Asian hockey, Pakistan dodge past India and reach the finals of #HockeyChampionsTrophy2014. Gre | Neutral | Negative | Negative |
| 8 | KiranKhanLive | RT @HafizZainAli: #defeat to #India in #India .WellDone team #Pakistan.#Hockey #ChampionsTrophy #ProudMoment | Negative | Neutral | Neutral |
| 9 | ZesHPirzadA | Talent hits a target no one else can hit & Genius hits the one no one else can see.that's why its 4-3 ;) #hockeychampionstrophy #PakvsInd | Positive | Neutral | Neutral |
| 10 | SportsReporter_ | BREAKING NEWS: Pakistan defeat India by 4-3 and reach the final of Champions Trophy Hockey.#PakvInd #CT2014 | Negative | Negative | Negative |
| 11 | therealfasih | RT @datelineAG: AG congratulates the Pakistan hockey team for winning the silver medal at the Champions Trophy. #GoPakistan! http://t.co/yn | Positive | Positive | Positive |
| 12 | zainabimam | Great, can we now move on to...#India's second defeat this week? RT @the hindu: 2 #Pakistan #hockey players suspended http://t.co/pXu4ghH544 | Negative | Positive | Positive |
| 13 | nadiakhaan | @Fizanism @ShkhRasheed @Insomniackhan91 Pakistani won Hockey match or Wrestling match. ... giving all different message! | Neutral | Positive | Positive |
| 14 | MuhamadAfzalECP | RT @Ali MuhammadPTI: Congratulations to Pakistan Hockey Team for a marvellous win against Arch Rivals India in India. | Positive | Negative | Positive |
| 15 | sharmeenochinoy | Well the commentary & the crowd are definitely not on #pakistan side! #hockey | Positive | Neutral | Neutral |
| 16 | ZesHPirzadA | RT @FIH Hockey: Magnificent eight for @EnglandHockey as they storm to biggest ever CT win over Pakistan #CT2014 #BestOfTheBest http://t.co | Negative | Negative | Negative |
| 17 | asmachaudhry24 | Well Done Pak Hockey team...beats India in India to earn a place in the final of Champions Trophy after 16 years...#GoTeamGreen | Negative | Negative | Positive |
| 18 | Asadullahk | RT @OnlineWorldTV: Congratulations to the Green Shirts for securing a tremendous victory! #PakvsInd #ChampionsTrophy #Hockey http://t.co/h | Positive | Negative | Positive |
| 19 | MurtazaGeoNews | Hero hockey captain thanks the whole team for playing their part in the historic win again India #PakvsInd http://t.co/tUNd1RJDYM | Positive | Positive | Positive |
| 20 | mak asif | @FIH Hockey - The Champions Trophy is considered the most prestigious & the toughest. Kindly keep it that why. | Positive | Neutral | Neutral |
| 21 | MurtazaGeoNews | RT @GeoJangPR: Twitter trend #ShamefulIndia shows "Pakistani nation is behind their hockey team & the players will play with all the prayer | Negative | Negative | Negative |
| 22 | ZahraPeer | Next time, please just make a V-sign with your fingers. It's easier for everybody. http://t.co/RxtVKttqPi #Pakstan #Hockey @FarazTalat | Positive | Positive | Positive |
| 23 | Khalil a hassan | RT @imranwaseem: shameless Govt has enough money to spend 10 millions /day for ads against IKBut wasnt paying few millions to hockey team | Negative | Negative | Positive |
| 24 | MehrTarar | RT @NadeemTameer: What a week for Pakistan! Youngest ever Nobel peace prize Pak born Amir Khan wins. Cricket team looking great. Hockey t | Negative | Positive | Positive |
| 25 | SaadiaAfzaal | RT @IjazANadeem: @SaadiaAfzaal MashaAllah, me too, so happy to learn here that Pakistan win in Hockey. Retrospecting my old days when Pak | Positive | Very Positive | Positive |
| 26 | adeelraja | Why is the Prime Minister praising Amir Khan the boxer only, whats wrong with the win of our hockey team win? | Negative | Negative | Negative |
| 27 | wajih sani | RT @omar quraishi: Classy of Indian media to deflect from Pak's brilliant win over the Indian hockey team by ad nauseum reporting on gestur | Positive | Negative | Negative |
| 28 | kashifsabir | RT @SalaamHockey: Congratulations to Germany for winning the Champions Trophy 2014! In other news, we're bringing a little bit of SILVER ba | Positive | Positive | Positive |
| 29 | hyzaidi | RT @KElectricPk: Wishing Team Green the best of luck for the big semi-final against India.#PakvsInd #Hockey #ChampionsTrophy http://t.co/I | Negative | Positive | Positive |
| 30 | usmanmanzoor | What a match of hockey, Great win by Pakistan | Positive | Positive | Positive |

(“ Pakistan authorities must ensure mil operations in N Waziristan respect laws of war, no collective punishment & provide for IDPs ”) from table 4.14, the journalists show their grievances about Zarb-e-Azab.

Figure 4.1b, shows the results of the event “Azadi and Inqlab March”. In this chart 8% of journalists’ opinions belong to a single category, 8% in two categories, 47% in three categories, 32% in four categories and 5% journalists’ opinions are detected in all the five categories.

In table 4.5, majority of tweets detected as negative in the event “Azadi and Inqlab March”. Actually in these tweets the most of the journalist criticized the government, election commission of Pakistan (ECP), and party chief of Jamiat Ulema-e-Islam. E.g. tweets from table 4.16 at serial no 1(“Maulana Fazlur Rehman makes terrible/incorrect accusation of fa’hashi against PTI’s protests.Stick to pol, Constitutional issues,Maulana sb”) and tweet at serial no 17 (“If army cannot help the people in getting justice, then it must step aside also & let the revolution takes its course !”) from table 4.8 support this event.

Figure 4.1c, shows the results of the event “Hockey Champions Trophy for Men 2014”. In this event 22% of journalist's opinions are categorized in a single category, 27% in two, 43% in three, 7% in four and only 1% of journalist's tweets belong to all the five categories.

From the results in table 4.5 it has been observed that most of the journalist express their opinion in negative sense because in the semifinal when Pakistani team celebrated its victory in the ground, International Hockey Federation (FIH) imposed one-match ban on two Pakistani players over misconducted and Pakistani Journalists criticized this act of FIH, e.g. in tweets “Whats wrong in our players jumping with joy and taking off their shirts?” and “Once again, shame on @FIH Hockey – #RoIndiaRo” shows the negative sense.

images/eventcluster.PNG

Figure 4.1: Sentiment Wise Journalists Cluster figure (a) for the event "Zarb-e-Azab", (b) for the event "Azadi and Inqlab March" and (c) for the event "Hockey Champions Trophy for Men 2014"

## 4.9 Discussion on Event Based Sentiments Classification

The sentiments analysis has drawback that, if any events have more negative tweets, it does not mean that people are talking against event, because sometimes there are circumstances, in which people are talking in the favor of the event by criticizing the opposition. To focus this problem, in this section we will discuss the sentiment results of each three event individually.

To clearly distinguish the sentiment of each event, we represent the results of table 4.5, in pie charts for each event separately in figure 4.2

As we discussed in section 4.4, that the event "Zarb-e-Azab" was against the militants. Figure 4.2a shows that there are 76% negative tweets. It means that majority of journalists were against the operation "Zarb-e-Azab". In tweets these journalists expressed their opinions regarding the issues in operation, militants, and problems faced by IDPS (internally displaced persons). Tweets at serial no 1, 3, 6, 7, 14, 16, and 17 in table 4.14 show that the majority of the journalists are not in favour of the operation. The tweets in table 4.14 are randomly selected from all the negative tweets of this event.

There are also 5% tweets, which are classified as positive. In the positive tweets, journalist are praising the operation and braveness of law enforcement agencies. For example tweets at serial no 1 and 7 in table 4.15 support our argument. The tweets in table 4.15 are randomly selected from the positive set.

Figure 4.2b shows the sentiment analysis of the event "Azadi and Inqlab March". As we already discussed in previous sections that this movement was against the government by two political parties PTI and PAT. As shown in the pie chart that 70% tweets are classified as negative, 22% as neutral and only 8% tweets are classified as positive. The majority of negative tweets do not mean that majority of journalists

images/sentimentBrakeup.PNG

Figure 4.2: Distribution of tweets by the sentiment classifier figure (a) for the event “Zarb-e-Azab”, (b) for the event “Azadi and Inqlab March” and (c) for the event “Hockey Champions Trophy for Men 2014”

Table 4.14: Tweets with Negative Sentiments from the Event “Zarb-e-Azab”

| S.No | Twitter id | Tweet | Polarity |
|---|---|---|---|
| 1 | AtifSal | #Pakistan rulers claims of #ZarbeAzb n #Waziristan exposed wth #PeshawarAttack.Fighting paid US War of Terror is bringing mayhem inside Pak. | Negative |
| 2 | KhSaad Rafique | Attended High level meeting chaired by Prime Minister. Current political situation , Operation , IDPs were discussed | Negative |
| 3 | MariumCh | RT @YusraSAskari: ’So far 572,529 people, belonging to 44,633 families have been registered as Internally Displaced Persons’ #ZarbEAzb #Pak | Negative |
| 4 | arsched | #PMSharif refused to authorise operation against militants when #GenSharif Gen sought authorisation in Feb & March #ZarbEAzb | Negative |
| 5 | DawarSafdar -  | RT @washingtonpost: Pakistan military advances against Taliban, kills 27 militants in North Waziristan http://t.co/3gMo14Rnbn | Negative |
| 6 | TahaSSiddiqui | #PakArmy commences op in #NorthWaziristan by name of #ZarbeAzb (Prophet’s sword name). Will it b a failure like Swat & South Waziristan ops? | Negative |
| 7 | Mustafa Qadri | Pakistan authorities must ensure mil operations in N Waziristan respect laws of war, no collective punishment & provide for IDPs | Negative |
| 8 | wasi78 | Terrorists’ network worth Rs 2 billion 49 crore 80 lakh destroyed as Zarb-e-Azb continues —... http://t.co/bYaDYGgk65 | Negative |
| 9 | taahir khan | Blast at marketplace in North Waziristan’s headquarters Miran-shah killed two people on Tuesday, tribesmen said. | Negative |
| 10 | PATofficialPK | #ZarbEAzb #ZarbeHaq rally by #PAT in Lahore to show solidarity with #PakArmy operation #Pakistan http://t.co/qgpa5i5SQk | Negative |
| 11 | MahsudFarooq | 2 soldiers wer killed,4 injd wen n IED planted by Terrorists, ex-ploded on roadside in area vill Jatarai barwand n South Waziris-tan Agency, | Negative |
| 12 | asmashirazi | RT @fareedraees: TTP’s Hafiz Gul Bahadur instructions to the people of #Waziristan. He has advised people to migrate before 10th June. http | Negative |

| | | | |
|---|---|---|---|
| 13 | ZaaraAbbasKhar | RT @AsimBajwaISPR: Army #Chief visited CMH,met injured students.Students said,we are in high spirits,consider us soldiers of ZarbeAzb,don't | Negative |
| 14 | shaistaAziz | How many people people remain in north Waziristan and which groups is the army targeting? http://t.co/AIiBjqUYCk via #Pakistan #ZarbeAzb | Negative |
| 15 | aliarqam | RT @mjdawar: Waziristan has been Razed to the ground by PAF Jets and none of the terrorists killed. #StateFranchisedTerror | Negative |
| 16 | DawarSafdar | Till curfew in waziristan,I think and worry that with curfew uplifting the war will start among military and militancy . | Negative |
| 17 | SanaTGulzar | RT @iramabbasi: Tribal customs making it difficult for some women IDPs to get access to all the help the need:My Video Story @BBCUrdu http: | Negative |
| 18 | arsched | RT @javerias: Chief of Army Staff General Raheel Sharif at MiranShah . #ZarbEAzb #MilataryOperation #NorthWazirstan #TTP #Pakistan http://t | Negative |
| 19 | Rabail26 | Extremist religious outfits have access to #IDPs in Bannu to provide "relief": by @TahaSSiddiqui. http://t.co/SbAGoPBYwu #ZarbeAzb #Pakistan | Negative |
| 20 | kazmiwajahat | BREAKING: #PakArmy troops are deployed in all the major cities of #Pakistan including #Karachi, #Lahore & #Islamabad. #ZarbeAzb #TTP | Negative |

Table 4.15: Tweets with Positive Sentiments from the Event “Zarb-e-Azab”

| S.No | Twitter id | Tweet | Polarity |
|---|---|---|---|
| 1 | khushnood2020 | RT @amarbail1: I am sure helpin IDPs in holy month of #Ramzan will bring peace and satisfaction to ur heart. #HelpIDPs @ErumManzoor http:// | Positive |
| 2 | MishalHusainBBC | Happy Christmas, wherever you are RT @Razarumi: Church in South Waziris-tan celebrates http://t.co/wK07ymz2 | Positive |
| 3 | FauziaKasuri | RT @syedsuhaibshah: Mrs. @FauziaKasuri distributing relief goods among the IDPs of North Waziristan with team@GVPakistan. @rameez mumtaz ht | Positive |
| 4 | sharmeenochinoy | Now would be a good time to address the nation! #PM #ZarbEAzb | Positive |
| 5 | NadiaaQasim | Plz Allah protect our soldiers who r in #WaziristanOperation as they always protect us.Allah Bless U and may you come home safe & sound,, | Positive |
| 6 | AtifSal | How many $$$?”@AsimBajwaISPR: #ZarbeAzb:Whole of nation approach will help us succeed vs terrorism,extremism in st http://t.co/VEQWUrZEHG” | Positive |
| 7 | taahir khan | Air strikes in Waziristan 'effective and successful': Sartaj Aziz http://t.co/9e81NV21el | Positive |
| 8 | TheHaroonRashid | A selfie with lovely Waziristan orphans at Sweet Home. Watch report now on @BBCurdu on Aaj TV http://t.co/QOzuwQnpB6 | Positive |
| 9 | omar quraishi | RT @Majid Agha: Dear @AsimBajwaISPR #ZarbEAzb is hope of the na-tion.#BringBackTaseerAndGillani http://t.co/kdUJ8cW2k9 | Positive |
| 10 | praveenswami | Jibran Ahmad has a great piece on refugees fleeing Pakistans war-torn North Waziristan — http://t.co/VyfkEskSIt | Positive |

were against the “Azadi and Inqlab March”. We randomly chose 20 negative tweets shown in table 4.16 to study why majority of the tweets are classified as negative. In the tweets at serial no 1, 6 and 20 journalists give opinion with negative sentiments about the said event and in tweet at serial no 10 comparing the march with previous movements and in other tweets peoples are expressing their support to march, with criticizing the government and others.

The random sample of tweets having positive sentiments is shown in table 4.17. In these tweets journalist express their opinion directly in the favour of the event. The examples of these type tweets are serial no 2, 3, 4, 8 and 9 in table 4.17.

Figure 4.2c shows the tweets classification of the event “Hockey Champions Trophy for Men 2014”. The tweets of this event are classified as 65% negative, 13% positive and 25% neutral. The more negative sentiments are because of the criticism on international hockey federation (FIH). FIH banned two Pakistani hockey players due to violence of discipline. For further analysis, we randomly choose 20 tweets as shown in table 4.18 from negatively classified tweets. In tweets at serial no 5, 6, 8, 15, 18 and 19 the people express their disgust about the decision of FIH, Indian media and

Table 4.16: Tweets with Negative Sentiments from the Event "Azadi and Inqlab March"

| S.No | Twitter id | Tweet | Polarity |
|---|---|---|---|
| 1 | NasimZehra | Maulana Fazlur Rehman makes terrible/incorrect accusation of fa'hashi against PTI's protests.Stick to pol, Constitutional issues,Maulana sb | Negative |
| 2 | Mahamali05 | Will the reforms include making one joint electorate for all? Will you ask for this reform Imran Khan? | Negative |
| 3 | ZaidZamanHamid | That is why Altaf the toad and beaten up, corrupt politicians are supporting Dr. TUQ. They want their share in the National govt ! Got it ? | Negative |
| 4 | MominaKhawar | Life. "@fasi zaka: A day after meeting his hero, Imran Khan's biggest fan passes away http://t.co/qdPnLuaNyX" | Negative |
| 5 | shabbeerwahgra | RT @saleemiss: Imran Khan releasing his workers from the Police Station, making a video & uploading it from his official page too https://t | Negative |
| 6 | ImaanZHazir | RT @mazdaki: After calling off his dharna Dr.Tahir-ul-Qadri walks straight into the dustbin of history; Aabpara will retrieve him if & when | Negative |
| 7 | mushtaqminhas | RT @Maria95PTI: @Asad Umar @ImranKhanPTI Almost many many pti girl wing ISB are upset ...We will not come in azadi march if its in allian | Negative |
| 8 | FaisalJavedKhan | VIDEO: Imran Khan's speech on 41st day of the protest at #AzadiSquare 23rd Sep, 2014 http://t.co/28coSjml5r | Negative |
| 9 | SikanderBalouch | @AnsarAAbbasi or Ulma Counsal ny wese bhe pehle bewi sy ijzat walee shart bhee khatm kardee,,ab tu Naya Pakistan with New Wife #AzadiSquare | Negative |
| 10 | tariqbutt | Particularly sacrifices of interior Sindh ppl in MRD movement were matchless Then many revolutionaries of 2day weren't even born in politics | Negative |
| 11 | KlasraRauf | RT @aslammuz: @Uzma Views @KlasraRauf @arsched media is also responsi- ble for this, promoting a criminal as a hero like @ImranKhanAnchor | Negative |
| 12 | AzazSyed | Met two doctors both support @ImranKhanPTI and both admit he has lack of vision. | Negative |
| 13 | sanabucha | Resignations? PTI in an effort to prove 'we mean business' by going 'out of business'! Only 'business as usual' in KPK. Vah! | Negative |
| 14 | SaeedShah | RT @MurtazaGeoNews: Leading female reporters @Fereeha and @asmashirazi threaten to boycott #PTI coverage if attacks on journalists by #PTI | Negative |
| 15 | mohsinrz | RT @KamranShafi46: 9,000 in one DHARNA and 11,000 in another ain't makin' Nawaz/Shahbaz Sharif to 'GO' and the assemblies to be dissolved.B | Negative |
| 16 | ShahidMursaleen | RT @RaheeqAbbasi: One of the reason for #DrQadri to go abroad: Govt denied right of treatment in Pakistan for #DrQadri | Negative |
| | | #LongLiveDrQadri htt | |
| 17 | Khalil a hassan | RT @TahirulQadri: We condemn the death of a protesting #PTI worker in<br>Faisalabad and the state brutality towards them. #PAT | Negative |
| 18 | wajih sani | RT @HamidMirGEO: Imran Khan mentioned missing persons after a long time<br>good to hear that from him | Negative |
| 19 | DrAwab | RT @ArsalanGhumman: Dharna has ended but uprising movement has stated<br>! #RespectForIK | Negative |
| 20 | kdastgirkhan | MT @AnsarAAbbasi:Breaking DI Khan & Bannu Jails was terrorism. Remov-<br>ing prisoners forcibly from police van was political activity? | Negative |

Table 4.17: Tweets with positive Sentiments from the Event “Azadi and Inqlab March”

| S.No | Twitter id | Tweet | Polarity |
|---|---|---|---|
| 1 | NazBalochPTI | Musarrat Misbah joins PTI. Great to have a leading woman in the field of social welfare become part of #PTIFamily. http://t.co/QE5aNGOgSo | Positive |
| 2 | Fereeha | Just finished a very interesting meeting with @TahirulQadri at his home. He categorically denied rumours of a deal. http://t.co/Hwii2BJubK | Positive |
| 3 | mosharrafzaidi | This is the finest piece you will read on Imran Khan’s Plan C. Even Insafians may like it (if you read to the end). http://t.co/ruybDf9lf1 | Positive |
| 4 | jasmeenmanzoor | RT @AmnaKhanPTI: And the hilarious moment when IK cleans his sweating with his kameez. Baqio k tou tissues hi nahi khatam hote | Positive |
| 5 | kazmiwajahat | An attempt to get celebrities like Shahid Afridi & Wasim Akram & in return a good number of crowd. #ShameOnIK #PTI http://t.co/7gaItCQ59G | Positive |
| 6 | NadeemMalikLive | @ImranKhanPTI talking with @nadeemmalik , right now @SAMAATV http://t.co/9Cfg9tgsFK | Positive |
| 7 | Mubashirlucman | @AsimBajwaISPR sir when will you gives us good news of Nawaz Sharif arrest? #AzadimarchPTI #InqilabmarchwithDrQadri #DrQadri | Positive |
| 8 | FarahnazZahidi | Thank you Sargodha! Massive! And well done #PTI for such an organzied jalsa. | Positive |
| 9 | arsched | The rich man stops laughing when the revolution comes. Quote #Revolution-March | Positive |
| 10 | NazBalochPTI | RT @syedarr: @RNYousuf @NazBalochPTI nice to meet the enthusiastic PTI couple .. Like My brother and sister .. #GoNawazGo http://t.co/zp | Positive |

others. These are the major reasons that majority of sentiments of this event are classified as negative.

## 4.10 Discussions

To know the opinion of targeted community about the events under study, we identified the members of community by using the snowball sampling technique. During this process it has been observed that irrelevant users were also fetched along with actual members. So in the snowball sampling the effective filtering mechanism is necessary as we did through a semi-manual process.

For knowing the opinions of journalists about events under study, we performed event based sentiment analysis of their tweets. Popular sentiment analyzers (such as Stanford CoreNLP) classified the tweets into very positive, positive, neutral, negative, and very negative.

It is difficult to completely rely on sentiments, in making the opinion about an event. Because the meaning of positive and negative is depend on context. For example in the event “Azadi and Inqlab March” journalists condemn the federal government and election commission of Pakistan for the rigging in general election 2013 and the way, the federal government handled the issue of protest and sit-in (dharna) by using the paramilitary forces. In the same way journalist also criticized the Punjab government due to the incident of model town, Lahore (section 4.4). Due to the condemning and criticism the large part of event related tweets are classified as negative. Upon the analysis of these negative tweets, we reached at the opinion that targeted community was supporting these protests in their tweets.

Table 4.18: Tweets with Negative Sentiments from the Event "Hockey Champions Trophy for Men 2014"

| S.No | Twitter id | Tweet | Polarity |
|---|---|---|---|
| 1 | JavedAzizKhan | Indian hockey chief announces ending ties with #Pakistan : TV reports ... WTH ? .. #Hockey #ChampionsTrophy | Negative |
| 2 | SaadiaAfzaal | RT @usmanmanzoor: When will Malik Riaz announce plots and cash for the poor Hockey players ??? #Waiting | Negative |
| 3 | alisalmanalvi | Some asses are set on fire... Sore losers. https://t.co/sh9w6CAkXC #ChampionsTrophy2014 #PakvInd #Hockey | Negative |
| 4 | shakirhusain | RT @faizanlakhani: BLAST FROM THE PAST: This is India's Prab- hjot Singh during WC2010, after Indian's loss to Argentina. cc : @FIH Hockey htt | Negative |
| 5 | AsmatullahNiazi | congratulations #Indian #Media for success in getting ban from #FIH is it not a biased decisions #fihockey ?????????? | Negative |
| 6 | ApaAlii | RT @SalaamHockey: It was only 5 members of the 7k crowd who'd said derogatory things & unfortunately got the better of our boys, spoiling t | Negative |
| 7 | IffatHasanRizvi | RT @Anujmanocha: @iffathasanrizvi @imvkohli cricket ka badla hockey me ! wah!! . see u in the world cup 2015 | Negative |
| 8 | MuhamadAfzalECP | RT @AQpk: After years of #Indian abuse and pettiness, finally someone from #Pakistan pays them in kind. #PakistanHockeyTeam #TitForTat #We | Negative |
| 9 | yasmeen 9 | RT @faizanlakhani: Nadeem Omar, the businessman who helped Pak- istan Hockey team financially, announces gold medals for the players of Pakis | Negative |
| 10 | ArifAlvi | Sorry! I opted out of @arsched ARY show because did not want to give up a double whammy show of cricket and hockey @KlasraRauf | Negative |
| 11 | SikanderRJ | Congrats Pakistan Hockey team on reaching the final of the #CT2014 beating India 4-3 | Negative |
| 12 | khalidkhan787 | pics: Muhammad Tousiq, left and Ammad Shakeel Butt per- forming Sajda after beating Netherlands in Hockey Quarterfinals http://t.co/WYttkzw4vg | Negative |
| 13 | AnsarAAbbasi | Report- Hockey India calls off bilateral series with Pakistan. #RoIndi- aRo #RoIndiaRo #RoIndiaRo #RoIndiaRo #RoIndiaRo #RoIndiaRo | Negative |
| 14 | AdilNajam | Congrats #Pakistan for #Silver in #Hockey #ChampionsTro- phy.But uneeded controversy bad for #SouthAsia + for #Hockey. | Negative |
| | | http://t.co/0zHxzYQ6Wq | |
| 15 | AqilSajjad | India’s behavior in hockey after yesterday’s game & how its been behav-<br>ing in cricket for the last few years. totally shameful | Negativ<br>e |
| 16 | khawajaNNInew<br>s | Report Decision PAK Player #22 Ali Amjad<br>https://t.co/7HToKfnPUu<br>#CT2014 #Bhubaneswar #fihockey” | Negativ<br>e |
| 17 | faizanlakhani | Nadeem Omar, the businessman who helped Pakistan Hockey team<br>financially, announces gold medals for the players of Pakistan team.<br>#CT2014 | Negativ<br>e |
| 18 | asadrana74 | RT @Khan Arsalan: That awkward moment when World’s Largest<br>Democracy cries over a Hockey Match Defeat..#RoIndiaRo<br>#CT2014<br>http://t.co/uQLj | Negativ<br>e |
| 19 | ApaAlii | Why has @FIH Hockey not made Youtube live streaming available in<br>England!? | Negativ<br>e |
| 20 | AQpk | #Pakistan remember: Our hockey team defeated in semifinals Champi-<br>ons Trophy becuz #India lobbied @FIH Hockey to wrongfully ban 2 key<br>players | Negativ<br>e |

While in “zarb-e-azb” event related tweets, those classified as negative, were really negative tweets. The community under study was reluctant that this war was not our, and they were worry regarding the displacement of the common residents of area, where the operation was started.

In event “Hockey Champions Trophy for Men 2014”, large part of relevant tweets also classified as negative. The Pakistani journalist community expressed their opinion in support of Pakistani players and team while condemning and criticizing the decision of ban by the International Hockey Federation (IHF) (section 4.9).

From the experiments and evaluation it has been observed that sentiment analysis i.e. Polarity of tweets is only helpful in primarily and rapid opinion making about any specific event. To know the complete opinion of community about specified event qualitative analysis of tweets is necessary.

## 4.11 Summary

In this chapter, we gave the overview of tools and libraries used in our experiments. Detailed discussion was carried out on experiments, dataset, and results of our study. We also discussed the sentiments of a targeted community in our study.

# Chapter 5

# Conclusions and Future Work

## 5.1 Conclusions

In this study, we discussed the importance and role of social media among various communities. We also highlighted the influential role of communities through the use of social media. We hypothesized that “Communities tweet on the events occurred in their surroundings” and “Communities express their opinions in the tweets on the events occurred in their surroundings”. For verifying our hypotheses, we proposed a generic framework, which was used to know the diversity of opinions and sentiments of a targeted community about a specific event.

We chose the Pakistani journalist community as our targeted community and three different events in Pakistani context. Members of Pakistani journalists community were identified and retrieved through our proposed framework by incorporating the snow ball sampling method. We crawled tweets posted by community members and identified tweets related to specific events. The tweets for specific events were cleaned. User ids and URLs were removed. Slang words and abbreviations were replaced with proper and complete words. Stemming and lemmatization was also applied on each tweet. The framework processed the cleaned tweets and fine-grained sentiments were detected. For the accuracy of sentiments detection results, we tested our dataset on Stanford CoreNLP analyzer and synesketch library. We found the results of Stanford CoreNLP analyzer more accurate than synesketch.

The results were evaluated quantitatively as well as qualitatively. The precision and recall of each event is calculated separately. The higher precision and recall values show that our proposed framework is effective and efficient in finding the event based sentiments of a targeted community.

This research is valuable and useful for different types of institutions in their own context. The government, opposition parties and other political parties are interested to know the sentiments of a community about specific events. They use the results in designing the public motivational tools and set the directions of their policies accordingly.

Further they address these communities accordingly. For example: Governments are interested in knowing the sentiments of specific community, before improving education, health, and justice systems.

Opposition parties use the sentiments of specific community about specific events to strengthen the check and balancing system over the government projects and policies.

During the election seasons, the political parties can check the sympathies of specific community towards political parties in advance through event based sentiment analysis with our proposed framework.

The brands and marketing companies can investigate community needs and demands of specified products for specified problems through the sentiment analysis. When they identify real needs of a community, they fulfill their needs and market their products in a better way.

This framework is also useful for foreign countries and agencies. They can monitor the event based sentiments of targeted communities of other foreign countries and use them for their own purposes like targeted aid and check and balance.

There are some drawbacks of our proposed framework. The drawback includes manually choosing seed users to identify the members of a targeted community. For selecting the seed users, the user must have some basic knowledge of community under study. If the user has no knowledge about targeted community, then there is the chance of in-correct selection of seed users. In the same way the users of the system may have basic background information of events under study. Which surely helps in selection of keywords for filtering relevant tweets from the dataset. It is difficult and time consuming job to remove irrelevant users of community members through a semi-manual process, if the size of the dataset is large.

As in our research we focused Pakistani journalist community and English is not their native language. So in their tweets, they usually make spelling and grammatical mistakes, and they use Urdu, Roman Urdu and other local languages in their tweets. The popular sentiment analyzers like Stanford CoreNLP do not produce accurate results because of spelling and grammatical mistakes and do not recognize tweets in Pakistani local languages. Due to this tweets other than English language are classified as neutral. However for achieving better accuracy, we try to correct English spelling mistakes to some extent by using the WordNet dictionary.

Because of the short length feature of tweet; twitter users usually use abbreviations and slang words in their tweets (section 2.2). The sentiment analyzer does not understand these abbreviations and slang words used in tweets, and does not achieve expected accuracy in sentiment analysis. Although our handmade customized slang and abbreviation dictionary (table 4.2) helps to replace the abbreviations and slang words with full form English word or phrases to some extent. But this is not a perfect solution, because sometimes the abbreviations and slang words are replaced with wrong words due to their ambiguity. For example in our study during the sentiment analysis of tweets pertaining to “ Azadi and Inqlab March”, we have observed that the abbreviation “PM” is used in multiple tweets. First when we replaced the abbreviations “PM” with its full form “Private Message” as in Netlingo acronyms dictionary. But, when we read some of the relevant tweets manually, we found that the abbreviation “PM” is for “Prime Minister”. Therefore the complete acronym dictionary in Pakistani context is necessary for achieving better results.

## 5.2 Future Work

The direction of future research includes the requirement of fully automatic filtering algorithm, which filters the relevant members of a targeted community from irrelevant members. As we observe in the experiments that twitter users kept the ids of miscellaneous users and targeted community users in the same twitter lists.

In our proposed framework the tweets other than English language are excluded, because the sentiment analyzers do not understand the local Pakistani languages, especially Urdu and Roman Urdu. So there should be provision of sentiment analysis facility of Urdu and Roman Urdu tweets in our proposed framework.

To verifying proposed framework, all the experiments are conducted using desktop applications. There is a need to develop a web based application of our proposed framework. So it becomes easily usable and beneficial for other relevant institutions.

# Bibliography

[1] Ebizmba.com, “Top 15 most popular social networking sites — july 2015,” 2015. [Online]. Available: http://www.ebizmba.com/articles/ social-networking-websites

[2] www.wegov project.eu, “Where egovernment meets the esociety — july 2015,” 2015. [Online]. Available: http://www.wegov-project.eu/

[3] A. Zubiaga, H. Ji, and K. Knight, “Curating and contextualizing twitter stories to assist with social newsgathering,” in *Proceedings of the 2013 International Conference on Intelligent User Interfaces*, ser. IUI ’13. New York, NY, USA: ACM, 2013, pp. 213–224. [Online]. Available: http://doi.acm.org/10.1145/2449396.2449424

[4] Jarwar, Muhammad Aslam, et al. "CommuniMents: A framework for detecting community based sentiments for events." International Journal on Semantic Web and Information Systems (IJSWIS) 13.2 (2017): 87-108.

[5] D. Zhao and M. B. Rosson, “How and why people twitter: The role that micro-blogging plays in informal communication at work,” in *Proceedings of the ACM 2009 International Conference on Supporting Group Work*, ser. GROUP

'09. New York, NY, USA: ACM, 2009, pp. 243–252. [Online]. Available: http://doi.acm.org/10.1145/1531674.1531710

[6] A. L. Washington, J. B. Thatcher, D. Morar, and K. LePrevost, "What is the correlation between twitter, polls and the popular vote in the 2012 presidential election?(correction)," in *Correction)(February 17, 2015). American Political Science Association 2013 Annual Meeting*, 2015.

[7] C. K. Coursaris, Y. Yun, and J. Sung, "Twitter users vs. quitters: A uses and gratifications and diffusion of innovations approach in understanding the role of mobility in microblogging," in *Proceedings of the 2010 Ninth International Conference on Mobile Business / 2010 Ninth Global Mobility Roundtable*, ser. ICMB-GMR '10. Washington, DC, USA: IEEE Computer Society, 2010, pp. 481–486. [Online]. Available: http://dx.doi.org/10.1109/ICMB-GMR.2010.44

[8] M. M. Mostafa, "More than words: Social networks' text mining for consumer brand sentiments," *Expert Syst. Appl.*, vol. 40, no. 10, pp. 4241–4251, Aug. 2013. [Online]. Available: http://dx.doi.org/10.1016/j.eswa.2013.01.019

[9] I. o. S. S. Blog Admin, "Available now: a guide to using twitter in university research, teaching, and impact activities," September 2011, © 2011 Impact of Social Sciences. [Online]. Available: http://eprints.lse.ac.uk/52098/

[10] B. O'Connor, R. Balasubramanyan, B. R. Routledge, and N. A. Smith, "From tweets to polls: Linking text sentiment to public opinion time series." *ICWSM*, vol. 11, no. 122-129, pp. 1–2, 2010.

[11] H. Kwak, C. Lee, H. Park, and S. Moon, "What is twitter, a social network or a news media?" in *Proceedings of the 19th International Conference on World Wide Web*, ser. WWW '10. New York, NY, USA: ACM, 2010, pp. 591–600. [Online]. Available: http://doi.acm.org/10.1145/1772690.1772751

[12] C. Honey and S. Herring, "Beyond microblogging: Conversation and collaboration via twitter," in *System Sciences, 2009. HICSS '09. 42nd Hawaii International Conference on*, Jan 2009, pp. 1–10.

[13] M. T. Bastos, C. Puschmann, and R. Travitzki, "Tweeting across hashtags: Overlapping users and the importance of language, topics, and politics," in *Proceedings of the 24th ACM Conference on Hypertext and Social Media*, ser. HT '13. New York, NY, USA: ACM, 2013, pp. 164–168. [Online]. Available: http://doi.acm.org/10.1145/2481492.2481510

[14] F. Kivran-Swaine, S. Brody, and M. Naaman, "Effects of gender and tie strength on twitter interactions," *First Monday*, vol. 18, no. 9, 2013.

[15] E. Riloff, J. Wiebe, and T. Wilson, "Learning subjective nouns using extraction pattern bootstrapping," in *Proceedings of the Seventh Conference on Natural Language Learning at HLT-NAACL 2003 - Volume 4*, ser. CONLL '03. Stroudsburg, PA, USA: Association for Computational Linguistics, 2003, pp. 25–32. [Online]. Available: http://dx.doi.org/10.3115/1119176.1119180

[16] S.-M. Kim and E. Hovy, "Determining the sentiment of opinions," in *Proceedings of the 20th International Conference on Computational Linguistics*, ser. COLING '04. Stroudsburg, PA, USA: Association for Computational Linguistics, 2004. [Online]. Available: http://dx.doi.org/10.3115/1220355.1220555

[17] B. Liu, "Sentiment analysis and opinion mining," *Synthesis Lectures on Human Language Technologies*, vol. 5, no. 1, pp. 1–167, 2012.

[18] W. James, "What is an emotion?" *Mind*, vol. 9, no. 34, pp. pp. 188–205, 1884. [Online]. Available: http://www.jstor.org/stable/2246769

[19] P. Ekman and W. V. Friesen, "Constants across cultures in the face and emotion." *Journal of personality and social psychology*, vol. 17, no. 2, p. 124, 1971.

[20] Jarwar, Muhammad Aslam, and Ilyoung Chong. "Exploiting IoT services by integrating emotion recognition in Web of Objects." 2017 International Conference on Information Networking (ICOIN). IEEE, 2017.

[21] G. Sidorov, S. Miranda-Jiménez, F. Viveros-Jiménez, A. Gelbukh, N. Castro-Sánchez, F. Velásquez, I. Díaz-Rangel, S. Suárez-Guerra, A. Treviño, and J. Gordon, “Empirical study of machine learning based approach for opinion mining in tweets,” in *Proceedings of the 11th Mexican International Conference on Advances in Artificial Intelligence - Volume Part I*, ser. MICAI’12. Berlin, Heidelberg: Springer-Verlag, 2013, pp. 1–14. [Online]. Available: http://dx.doi.org/10.1007/978-3-642-37807-2 1

[22] M. Tsytsarau, T. Palpanas, and K. Denecke, “Scalable detection of sentiment-based contradictions,” in *DiversiWeb 2011: First International Workshop on Knowledge Diversity on the Web, in conjunction with WWW 2011, March 28, 2011, Hyderabad*, 2011.

[23] S. Wen and X. Wan, “Emotion classification in microblog texts using class sequential rules,” in *Twenty-Eighth AAAI Conference on Artificial Intelligence, held on July 27 to 31 at Qubec Convention Center, Qubec City, Qubec, Canada*, 2014.

[24] T. Nasukawa and J. Yi, “Sentiment analysis: Capturing favorability using natural language processing,” in *Proceedings of the 2Nd International Conference on Knowledge Capture*, ser. K-CAP ’03. New York, NY, USA: ACM, 2003, pp. 70–77. [Online]. Available: http://doi.acm.org/10.1145/945645.945658

[25] W. Medhat, A. Hassan, and H. Korashy, “Sentiment analysis algorithms and applications: A survey,” *Ain Shams Engineering Journal*, vol. 5, no. 4, pp.

1093 – 1113, 2014. [Online]. Available: http://www.sciencedirect.com/science/article/pii/S2090447914000550

[26] P. D. Turney, "Thumbs up or thumbs down?: Semantic orientation applied to unsupervised classification of reviews," in *Proceedings of the 40th Annual Meeting on Association for Computational Linguistics*, ser. ACL '02. Stroudsburg, PA, USA: Association for Computational Linguistics, 2002, pp. 417–424. [Online]. Available: http://dx.doi.org/10.3115/1073083.1073153

[27] B. Pang and L. Lee, "Opinion mining and sentiment analysis," *Found. Trends Inf. Retr.*, vol. 2, no. 1-2, pp. 1–135, Jan. 2008. [Online]. Available: http://dx.doi.org/10.1561/1500000011

[28] T. Wilson, J. Wiebe, and P. Hoffmann, "Recognizing contextual polarity in phrase-level sentiment analysis," in *Proceedings of the Conference on Human Language Technology and Empirical Methods in Natural Language Processing*, ser. HLT '05. Stroudsburg, PA, USA: Association for Computational Linguistics, 2005, pp. 347–354. [Online]. Available: http://dx.doi.org/10.3115/1220575.1220619

[29] A. Pak and P. Paroubek, "Twitter as a corpus for sentiment analysis and opinion mining." in *Language Resources and Evaluation (LREC)*, vol. 10, 2010, pp. 1320–1326.

[30] M. Hu and B. Liu, "Mining and summarizing customer reviews," in *Proceedings of the Tenth ACM SIGKDD International Conference on Knowledge Discovery and Data Mining*, ser. KDD '04. New York, NY, USA: ACM, 2004, pp. 168–177. [Online]. Available: http://doi.acm.org/10.1145/1014052.1014073

[31] P. Burnap, R. Gibson, L. Sloan, R. Southern, and M. L. Williams, "140 characters to victory?: Using twitter to predict the UK 2015

general election," *CoRR*, vol. abs/1505.01511, 2015. [Online]. Available: http://arxiv.org/abs/1505.01511

[32] A. Tsakalidis, S. Papadopoulos, A. Cristea, and Y. Kompatsiaris, "Predicting elections for multiple countries using twitter and polls," *Intelligent Systems, IEEE*, vol. 30, no. 2, pp. 10–17, Mar 2015.

[33] A. Bermingham and A. Smeaton, "On using twitter to monitor political sentiment and predict election results," in *Proceedings of the Workshop on Sentiment Analysis where AI meets Psychology (SAAIP 2011)*. Chiang Mai, Thailand: Asian Federation of Natural Language Processing, November 2011, pp. 2–10. [Online]. Available: http://www.aclweb.org/anthology/W11-3702

[34] Y. Liu, X. Huang, A. An, and X. Yu, "Arsa: A sentiment-aware model for predicting sales performance using blogs," in *Proceedings of the 30th Annual International ACM SIGIR Conference on Research and Development in Information Retrieval*, ser. SIGIR '07. New York, NY, USA: ACM, 2007, pp. 607–614. [Online]. Available: http://doi.acm.org/10.1145/1277741.1277845

[35] J. Bollen, H. Mao, and X. Zeng, "Twitter mood predicts the stock market," *Journal of Computational Science*, vol. 2, no. 1, pp. 1 – 8, 2011. [Online]. Available: http://www.sciencedirect.com/science/article/pii/S187775031100007X

[36] C. C. Aggarwal and C. X. Zhai, *Mining Text Data*. Springer Publishing Company, Incorporated, 2012.

[37] F. H. Khan, S. Bashir, and U. Qamar, "Tom: Twitter opinion mining framework using hybrid classification scheme," *Decis. Support Syst.*, vol. 57, pp. 245–257, Jan. 2014. [Online]. Available: http://dx.doi.org/10.1016/j.dss.2013.09.004

[38] S. Tan and J. Zhang, “An empirical study of sentiment analysis for chinese documents,” *Expert Syst. Appl.*, vol. 34, no. 4, pp. 2622–2629, May 2008. [Online]. Available: http://dx.doi.org/10.1016/j.eswa.2007.05.028

[39] C. D. Manning, M. Surdeanu, J. Bauer, J. Finkel, S. J. Bethard, and D. McClosky, “The stanford corenlp natural language processing toolkit,” in *Proceedings of 52nd Annual Meeting of the Association for Computational Linguistics: System Demonstrations*, 2014, pp. 55–60.

[40] U. Krcadinac, P. Pasquier, J. Jovanovic, and V. Devedzic, “Synesketch: An open source library for sentence-based emotion recognition,” *Affective Computing, IEEE Transactions on*, vol. 4, no. 3, pp. 312–325, July 2013.

[41] www.ling.upenn.edu, “Alphabetical list of part-of-speech tags used in the penn treebank project — july 2015,” 2015. [Online]. Available: http://www.ling.upenn.edu/courses/Fall_2003/ling001/penn_treebank_pos.html

[42] M.-C. de Marneffe and C. D. Manning, “The stanford typed dependencies representation,” in *Coling 2008: Proceedings of the Workshop on Cross-Framework and Cross-Domain Parser Evaluation*, ser. CrossParser ’08. Stroudsburg, PA, USA: Association for Computational Linguistics, 2008, pp. 1–8. [Online]. Available: http://dl.acm.org/citation.cfm?id=1608858.1608859

[43] R. Socher, A. Perelygin, J. Wu, J. Chuang, C. D. Manning, A. Ng, and C. Potts, “Recursive Deep Models for Semantic Compositionality Over a Sentiment Treebank,” in *Proceedings of the 2013 Conference on Empirical Methods in Natural Language Processing*. Seattle, Washington, USA: Association for Computational Linguistics, Oct. 2013, pp. 1631–1642. [Online]. Available: http://www.aclweb.org/anthology-new/D/D13/D13-1170.bib

[44] S. Wu, J. M. Hofman, W. A. Mason, and D. J. Watts, “Who says what to whom on twitter,” in *Proceedings of the 20th International Conference on World Wide Web*, ser. WWW ’11. New York, NY, USA: ACM, 2011, pp. 705–714. [Online]. Available: http://doi.acm.org/10.1145/1963405.1963504

[45] W. Li and H. Xu, “Text-based emotion classification using emotion cause extraction,” *Expert Syst. Appl.*, vol. 41, no. 4, pp. 1742–1749, Mar. 2014. [Online]. Available: http://dx.doi.org/10.1016/j.eswa.2013.08.073

[46] P. Burnap, O. F. Rana, N. Avis, M. Williams, W. Housley, A. Edwards, J. Morgan, and L. Sloan, “Detecting tension in online communities with computational twitter analysis,” *Technological Forecasting and Social Change*, vol. 0, no. 0, pp. –, 2013, -. [Online]. Available: http://www.sciencedirect.com/science/article/pii/S0040162513000899

[47] R. Kempter, V. Sintsova, C. C. Musat, and P. Pu, “Emotionwatch: Visualizing fine-grained emotions in event-related tweets,” in *Proceedings of the Eighth International Conference on Weblogs and Social Media, ICWSM 2014, Ann Arbor, Michigan, USA, June 1-4, 2014.*, 2014. [Online]. Available: http://www.aaai.org/ocs/index.php/ICWSM/ICWSM14/paper/view/8117

[48] Y. Bao, C. Quan, L. Wang, and F. Ren, “The role of pre-processing in twitter sentiment analysis,” in *Intelligent Computing Methodologies*, ser. Lecture Notes in Computer Science, D.-S. Huang, K.-H. Jo, and L. Wang, Eds. Springer International Publishing, 2014, vol. 8589, pp. 615–624. [Online]. Available: http://dx.doi.org/10.1007/978-3-319-09339-0_62